\documentclass[twocolumn,prl,floatfix,superscriptaddress,nobibnotes]{revtex4-1}

\usepackage{amsmath}
\usepackage{amssymb}
\usepackage{graphicx}
\usepackage{hyperref}
\hypersetup{colorlinks=true,linkcolor=black,citecolor=black}
\usepackage{color}

\begin{document}
\title{Supercurrent reversal in ferromagnetic hybrid nanowire Josephson junctions}
\author{D.~Razmadze}
\thanks{These two authors contributed equally.}
\affiliation{Center for Quantum Devices, Niels Bohr Institute, University of Copenhagen, 2100 Copenhagen, Denmark}
\author{R.~Seoane~Souto}
\thanks{These two authors contributed equally.}
\affiliation{Center for Quantum Devices, Niels Bohr Institute, University of Copenhagen, 2100 Copenhagen, Denmark}
\affiliation{Division of Solid State Physics and NanoLund, Lund University, 22100 Lund, Sweden}
\author{L.~Galletti}
\affiliation{Center for Quantum Devices, Niels Bohr Institute, University of Copenhagen, 2100 Copenhagen, Denmark}
\author{A.~Maiani}
\affiliation{Center for Quantum Devices, Niels Bohr Institute, University of Copenhagen, 2100 Copenhagen, Denmark}
\author{Y.~Liu}
\affiliation{Center for Quantum Devices, Niels Bohr Institute, University of Copenhagen, 2100 Copenhagen, Denmark}
\author{P.~Krogstrup}
\affiliation{Center for Quantum Devices, Niels Bohr Institute, University of Copenhagen, 2100 Copenhagen, Denmark}
\author{C.~Schrade}
\affiliation{Center for Quantum Devices, Niels Bohr Institute, University of Copenhagen, 2100 Copenhagen, Denmark}
\author{A.~Gyenis}
\thanks{Current address: Department of Electrical, Computer \& Energy Engineering, University of Colorado Boulder, CO 80309, USA}
\affiliation{Center for Quantum Devices, Niels Bohr Institute, University of Copenhagen, 2100 Copenhagen, Denmark}
\author{C.~M.~Marcus}
\affiliation{Center for Quantum Devices, Niels Bohr Institute, University of Copenhagen, 2100 Copenhagen, Denmark}
\author{S.~Vaitiek\.{e}nas}
\affiliation{Center for Quantum Devices, Niels Bohr Institute, University of Copenhagen, 2100 Copenhagen, Denmark}
\date{\today}


\begin{abstract}
We report supercurrent transport measurements in hybrid Josephson junctions comprised of semiconducting InAs nanowires with epitaxial ferromagnetic insulator EuS and superconducting Al coatings.
The wires display a hysteretic superconducting window close to the coercivity, away from zero external magnetic field.
Using a multi-interferometer setup, we measure the current-phase relation of multiple magnetic junctions and find an abrupt switch between $\pi$ and 0 phases within the superconducting window.
We attribute the 0\,--\,$\pi$ transition to the discrete flipping of the EuS domains and provide a qualitative theory showing that a sizable exchange field can polarize the junction and lead to the supercurrent reversal.
Both $0$ and $\pi$ phases can be realized at zero external field by demagnetizing the wire.
\end{abstract}

\maketitle


Unexpected quantum behavior often appears in hybrid devices made from materials with competing electrical properties~\cite{Meservey1994, Bergeret2018, Prada2020}.
For instance, Josephson junctions with ferromagnetic barriers can develop a superconducting phase difference of $\pi$ giving rise to supercurrent reversal~\cite{Ryazanov2001, Kontos2002} and spontaneous currents in superconducting loops~\cite{Bauer2004}.
The superconducting phase shift in metallic samples typically originates from the oscillatory damping of the pairing amplitude in the junction~\cite{Buzdin2005, Bergeret2005}. However, it can also arise from spin-flip scattering via magnetic impurities  ~\cite{Bulaevskii1977,Schrade2015} or an alternating magnetization in multidomain junctions~\cite{Volkov2005}.
The dependence of the superconducting parameters on the magnetic domain structure has been investigated in samples with metallic~\cite{Yang2004} and insulating~\cite{Strambini2017, Diesch2018} ferromagnetic components, but the impact of domain configuration on the $\pi$-junction formation remains largely unexplored.

Spin-active scattering at an interface between a superconductor and a ferromagnetic insulator can induce a Zeeman-like splitting~\cite{Tokuyasu1988, Hao1990} and give rise to spin-polarized Andreev bound states (ABSs)~\cite{Beckmann2018}.
Superconducting junctions with large interfacial exchange interaction are predicted to develop $\pi$ phase shift~\cite{Tanaka1997, Fogelstrom2000, Minutillo2021} with intrinsically low quasiparticle dissipation~\cite{Kawabata2006,Kato2007}, making them attractive for both classical and quantum applications~\cite{Ioffe1999, Feofanov2010, Gingrich2016}. 
Previously, supercurrent transport through spin-dependent barriers showed signatures of unconventional superconductivity~\cite{Senapati2011} and incomplete 0\,--\,$\pi$ transition~\cite{Caruso2019} associated with the multidomain structure in the junction.

An alternative route to Josephson $\pi$-junctions is to couple a semiconducting quantum dot~\cite{Spivak1991,vanDam2006, Bouman2020} or superconducting hybrid island~\cite{Schrade2018,Wang2021} to superconducting leads, where adding or removing a single electron spin can result in a supercurrent reversal.
The transition between the 0 and $\pi$ phases in such junctions can be controlled electrostatically~\cite{Razmadze2020, Awoga2019, Schulenborg2020} or by an external magnetic field~\cite{Whiticar2021, Bargerbos2022}.
Recently developed semiconducting InAs nanowires with epitaxial superconducting Al and ferromagnetic insulator EuS shells~\cite{Krogstrup2015,Liu2019,Liu2019_2} showed coexistence of induced superconductivity and ferromagnetism, and signatures of spin-polarized bound states~\cite{Vaitiekenas2020, Vaitiekenas2022}.
The latter were investigated theoretically in the context of topological superconductivity~\cite{Woods2021, Maiani2021, Escribano2021, Liu2021, Langbehn2021, Khindanov2021, Liu2022, Escribano2022}.
Here, we study triple-hybrid Josephson junctions containing components of spin-dependent transport and gate-voltage controlled barriers.

\begin{figure}[t!]
\includegraphics[width=\linewidth]{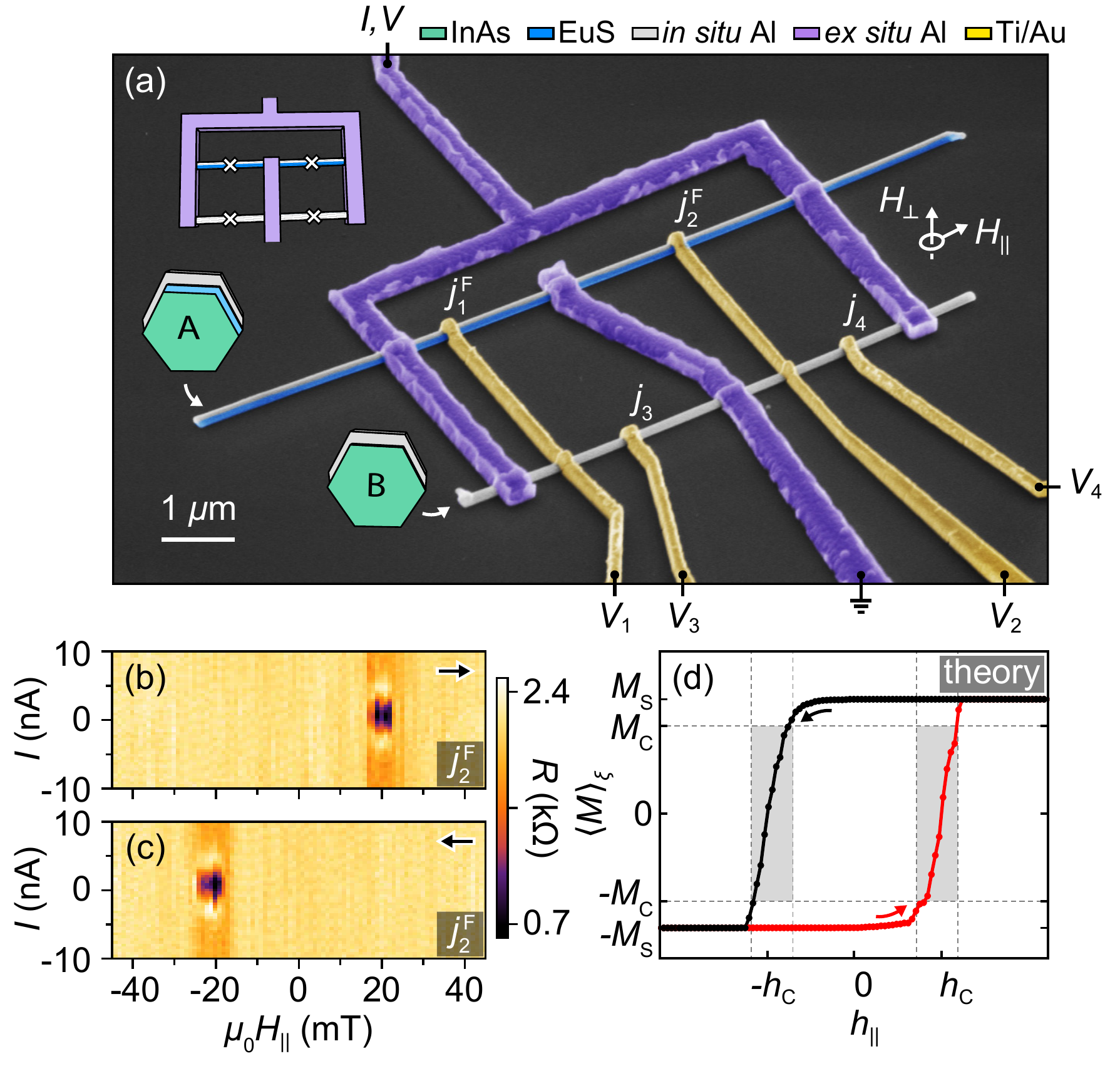}
\caption{\label{fig:1}
(a) Color-enhanced micrograph of a representative multi-interferometer device comprised of ferromagnetic (target) and non-magnetic (reference) nanowires. The insets show schematic device layout and wire cross-sections.
\mbox{(b) and (c)} Differential resistance, $R$, as a function of current bias, $I$, and parallel external magnetic field, $H_\parallel$, measured for the $j_2^{\rm F}$ junction sweeping $H_\parallel$ from (b) negative to positive and (c) positive to negative.
$R$ is suppressed in a narrow, sweep-direction dependent window away from $H_\parallel = 0$.
(d) Disorder-averaged induced magnetization, $\left< M \right>_\xi$, calculated using random-field Ising model.  $h_\parallel$ is a model parameter representing external magnetic field. The junction superconducts only in a narrow hysteretic window (gray) around the coercive field $\pm h_{\rm C}$ and is otherwise normal.
}
\end{figure}

To demonstrate the emergence of $\pi$-junctions in ferromagnetic hybrid nanowires, we studied a multi-interferometer device consisting of ferromagnetic (target) and non-magnetic (reference) wires.
The two wires, denoted A and B, were placed next to each other on a Si substrate with 200~nm SiOx capping.
The middle and the ends of both wires were connected by \textit{ex situ} Al contacts, forming multiple loops~[Fig.~\ref{fig:1}(a)].
The main wire~A was comprised of a hexagonal InAs core with epitaxial two-facet EuS and three-facet \textit{in~situ} Al shells, with the Al fully covering both EuS facets and one InAs facet.
The reference wire B with InAs core and three-facet Al shell did not include EuS.
Four junctions, denoted $j_1^{\rm F}$, $j_2^{\rm F}$ on the ferromagnetic wire A, and $j_3$, $j_4$ on the reference wire B, were formed by selectively removing $\sim100$~nm of \textit{in situ} Al in the segments between the ohmic contacts.
Top gates were fabricated over all four junctions after the deposition of a thin HfOx dielectric layer, allowing independent electrostatic control of each junction.

The phase across a particular junction relative to a reference junction was measured by depleting the other two junctions, thus forming a single superconducting interferometer.
Three triple-hybrid junctions, from two different devices, were investigated and showed similar results. We report data from a representative junction in the main text and present supporting data from the other two junctions in the Supplemental Material~\cite{Supplement}.
Measurements were carried out using standard ac lock-in techniques in a dilution refrigerator with a three-axis vector magnet and base temperature of 20 mK.

\begin{figure}[t!]
\includegraphics[width=\linewidth]{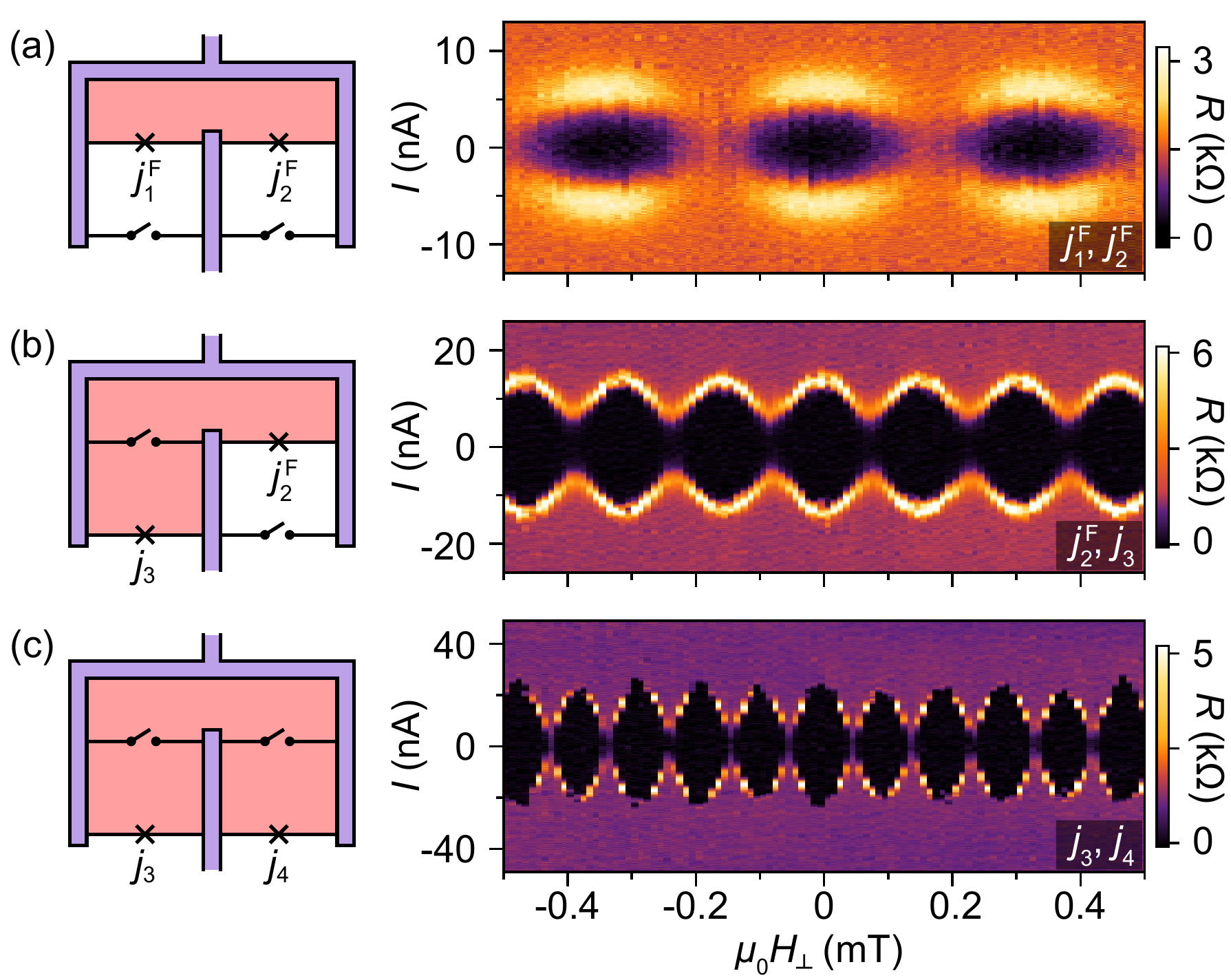}
\caption{\label{fig:2}
Left panels: Schematics of the multi-interferometer device in various open and closed junction configurations with highlighted effective loop areas.
Right panels: Corresponding current-phase relations represented by differential resistance, $R$, measured as a function of current bias, $I$, and flux-threading perpendicular magnetic field, $H_\perp$, for (a) two ferromagnetic, (b) one ferromagnetic and one non-magnetic, and (c) two non-magnetic junctions.
All junction configurations show effective-area dependent, periodic switching current modulations in $H_\perp$. 
}
\end{figure}

We begin by exploring magnetotransport properties of a single ferromagnetic junction, $j_2^{\rm F}$, while keeping the other junctions depleted.
Four-terminal differential resistance, $R = {\rm d}V/{\rm d}I$, of the junction was measured as a function of current bias, $I$, and external magnetic field, $H_\parallel$, applied parallel to wire A [Figs.~\ref{fig:1}(b) and \ref{fig:1}(c)].
Sweeping from negative to positive field, $R(I)$ remains finite and featureless throughout the measured range, except between $\mu_0 H_\parallel = 15$ and $25$~mT, where $R(I)$ decreases abruptly for $\vert I\vert\lesssim 5$~nA [Fig.~\ref{fig:1}(b)].
Reversing the sweep direction of $H_\parallel$ shifts the low-resistance window to around $-20$~mT [Fig.~\ref{fig:1}(c)].
A similar hysteretic dip in resistance has been reported in uninterrupted EuS/Al bilayer films~\cite{Li2013}.

We interpret the observed behavior as the recovery of the superconductivity near the coercive field, $H_{\rm C}$, where the induced magnetization, $\left<M\right>_\xi$, averaged over the superconducting coherence length, $\xi$, decreases below a critical value $M_{\rm C}$.
To verify this picture, we calculate disorder-averaged $\left<M\right>_\xi$ using the kinetic random-field Ising model~(see Supplemental Material~\cite{Supplement} and Refs.~\cite{Sethna1993,Glauber1963} therein).
The resulting hysteresis curves [see Fig.~\ref{fig:1}(d)] are asymmetric around the coercive field.
In this regime the EuS domain size is shorter than $\xi$, which leads to a reduced $\left<M\right>_\xi$ compared to the saturation value, $M_{\rm S}$ (see Supplemental Material~\cite{Supplement} and Refs.~\cite{Kittel2005,Tinkham1996,Vaitiekenas2019,Bruno1973,Chandrasekhar1962,Clogston1962} therein).
Realistic hysteresis curves are typically not as smooth as depicted in Fig.~\ref{fig:1}(d); instead, they display irreversible jumps between discrete magnetization values~\cite{Wolf2014,Vaitiekenas2020,Hijano2021}.

We note that the magnetic junctions in the superconducting state display residual resistance at $I=0$, which we tentatively attribute to the supercurrent suppression due to the uncertainty in the phase difference across a junction with low Josephson energy~\cite{Lu2021}. Such phase diffusion can be stabilized by integrating the junction into a superconducting loop~\cite{Sullivan2013}.

Having established the magnetic-field response of an individual magnetic junction, we next examine the current-phase relations (CPRs) of various junction pairs.
To ensure that wire~A was superconducting, $\mu_0 H_\parallel$ was first ramped to $-100$~mT and then tuned to $21$~mT, close to $H_{\rm C}$, where $R(I\sim0)$ is suppressed.
Three example measurements of distinct superconducting interferometers, formed by opening either two ferromagnetic, mixed, or non-magnetic junctions, are displayed in Fig.~\ref{fig:2}.
In all three configurations, the device shows periodic switching current, $I_{\rm SW}$, modulations as a function of the flux-threading perpendicular magnetic field, $H_\perp$.
The oscillation period changes for different junction combinations due to the different effective loop areas, corresponding to the superconducting flux quantum, $\Phi_0 = h/2e$ (Fig.~S1 in Supplemental Material~\cite{Supplement}).
The zero-flux offset of the magnet was calibrated using the CPR of the loop with two non-magnetic junctions ($j_3$ and $j_4$).

\begin{figure}[t!]
\includegraphics[width=\linewidth]{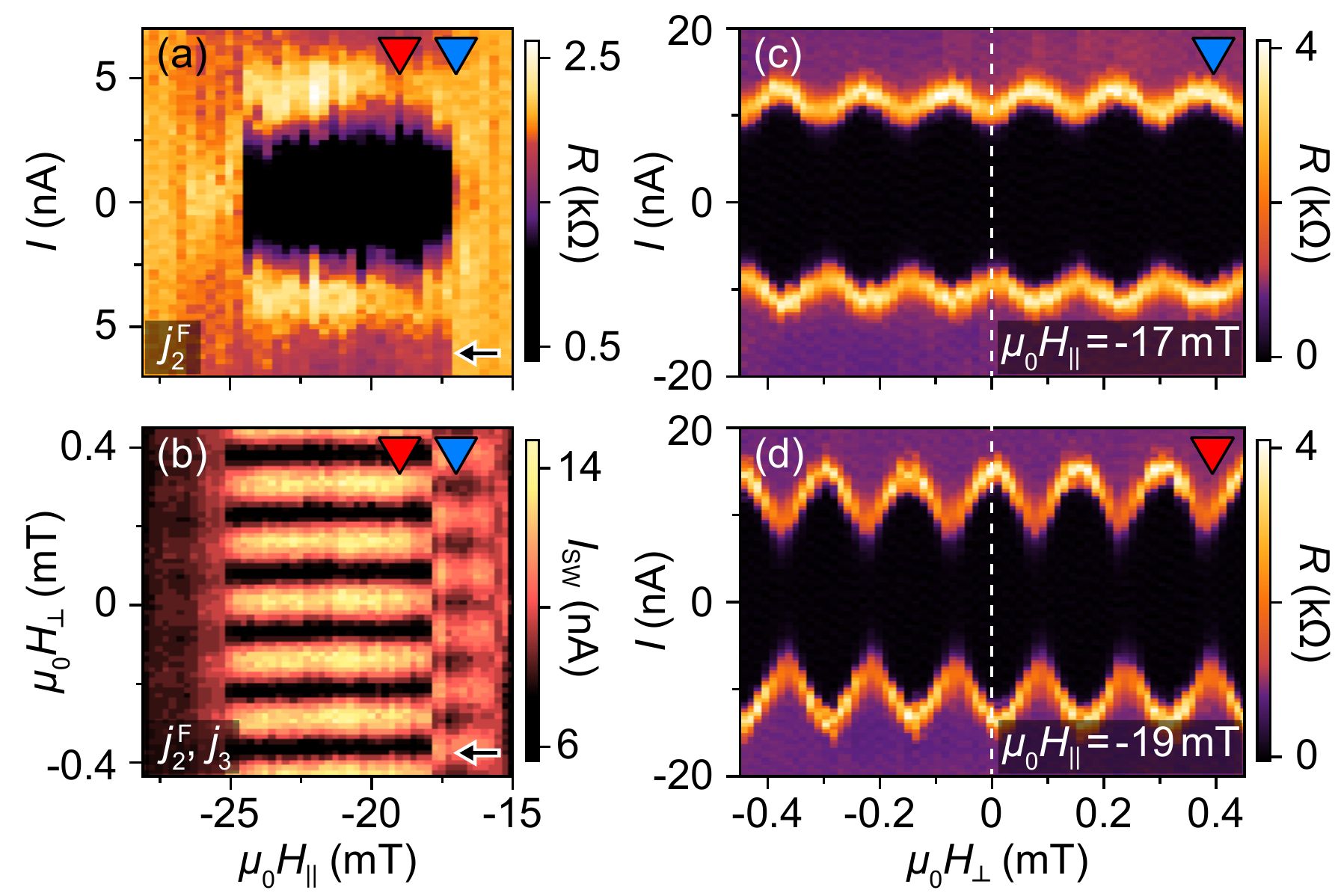}
\caption{\label{fig:3} 
(a) Differential resistance, $R$, of $j_2^{\rm F}$ as a function of current bias, $I$, and parallel external magnetic field, $H_\parallel$, showing the superconducting window of the junction centered around $-21$~mT.
The data were taken with $j_1^{\rm F}$, $j_3$, and $j_4$ depleted.
(b) Switching current, $I_{\rm SW}$, as a function of flux-threading perpendicular magnetic field, $H_\perp$, measured for $j_2^{\rm F}$\,-\,$j_3$ interferometer at decreasing $H_\parallel$ values.
The magnetic junction switches abruptly from $\pi$- to 0-phase around $-18$~mT as $H_\parallel$ is lowered.
(c) Current-phase relation measured at $\mu_0H_\parallel = -17$~mT exhibits $I_{\rm SW}$ minimum at $H_\perp = 0$, suggesting $\pi$-junction.
(d) similar to (c) but in the 0-junction regime at $\mu_0H_\parallel = -19$~mT.
All the data were taken after polarizing the wire at $\mu_0 H_\parallel = 100$~mT.
}
\end{figure}

At $\mu_0 H_\parallel = 21$~mT, $I_{\rm SW}$ is maximal at zero flux ($H_\perp = 0$) for all configurations, indicating a homogeneous superconducting phase across the device.
However, we find that the loops with magnetic junctions show characteristic $\pi$-shifted CPRs at the onset of the superconducting window, before $H_{\rm C}$ is reached~(Fig.~\ref{fig:3}).
We study the transition between the two regimes in $j_2^{\rm F}$ by measuring CPRs (using $j_3$ as a reference, while keeping $j_1^{\rm F}$ and $j_4$ depleted) over a range of $H_\parallel$ spanning the superconducting window~[Fig.~\ref{fig:3}(a)].
The deduced evolution of $I_{\rm SW}$ with $H_\perp$ and $H_\parallel$ is shown in Fig.~\ref{fig:3}(b).
Outside the superconducting window, $I_{\rm SW}$ is independent of $H_\perp$, indicating that $j_2^{\rm F}$ is not superconducting (Fig.~S2 in Supplemental Material~\cite{Supplement}).
Moving to more negative $H_\parallel$, $j_2^{\rm F}$ displays a $\pi$-shifted CPR in the range between $-16$ and $-18$~mT~[Fig.~\ref{fig:3}(c)], but then switches abruptly to a state without a phase shift~[Fig.~\ref{fig:3}(d)].
The average $I_{\rm SW}$ is $\sim 10$~nA in both cases, but its modulation amplitude increases from around 3 to 6~nA as the CPR phase switches from $\pi$ to 0 [see Figs.~\ref{fig:3}(c) and \ref{fig:3}(d)].
The superconducting phase remains unchanged throughout the rest of the superconducting window, whereas the amplitude of $I_{\rm SW}$ oscillations shrinks abruptly at $-25$~mT and once again at $-26$~mT as the supercurrent through $j_2^{\rm F}$ gets suppressed.
This is likely because of the sweep-direction dependent, discrete jumps of $\left<M\right>_\xi$ through $M_{\rm C}$ (see Supplemental Material~\cite{Supplement}).
The transition features were qualitatively the same around positive $H_{\rm C}$, after reversing $H_\parallel$ direction, and did not depend on the gate voltages $V_2$ and $V_3$ (see Figs.~S3, S4, and S5 in Supplemental Material~\cite{Supplement}).
The transition-field value shifted by a fraction of a millitesla between different runs, presumably due to the magnetic noise from the stochastic domain switching~\cite{Puppin2000}.
Furthermore, the superconducting phase shift of $\pi$ within the superconducting window was observed also for the other two measured ferromagnetic junctions, with $j_1^{\rm F}$ showing hints of a second 0\,--$\pi$ transition at the end of its superconducting window (see Figs.~S6 and S7).


\begin{figure}[t!]
\includegraphics[width=\linewidth]{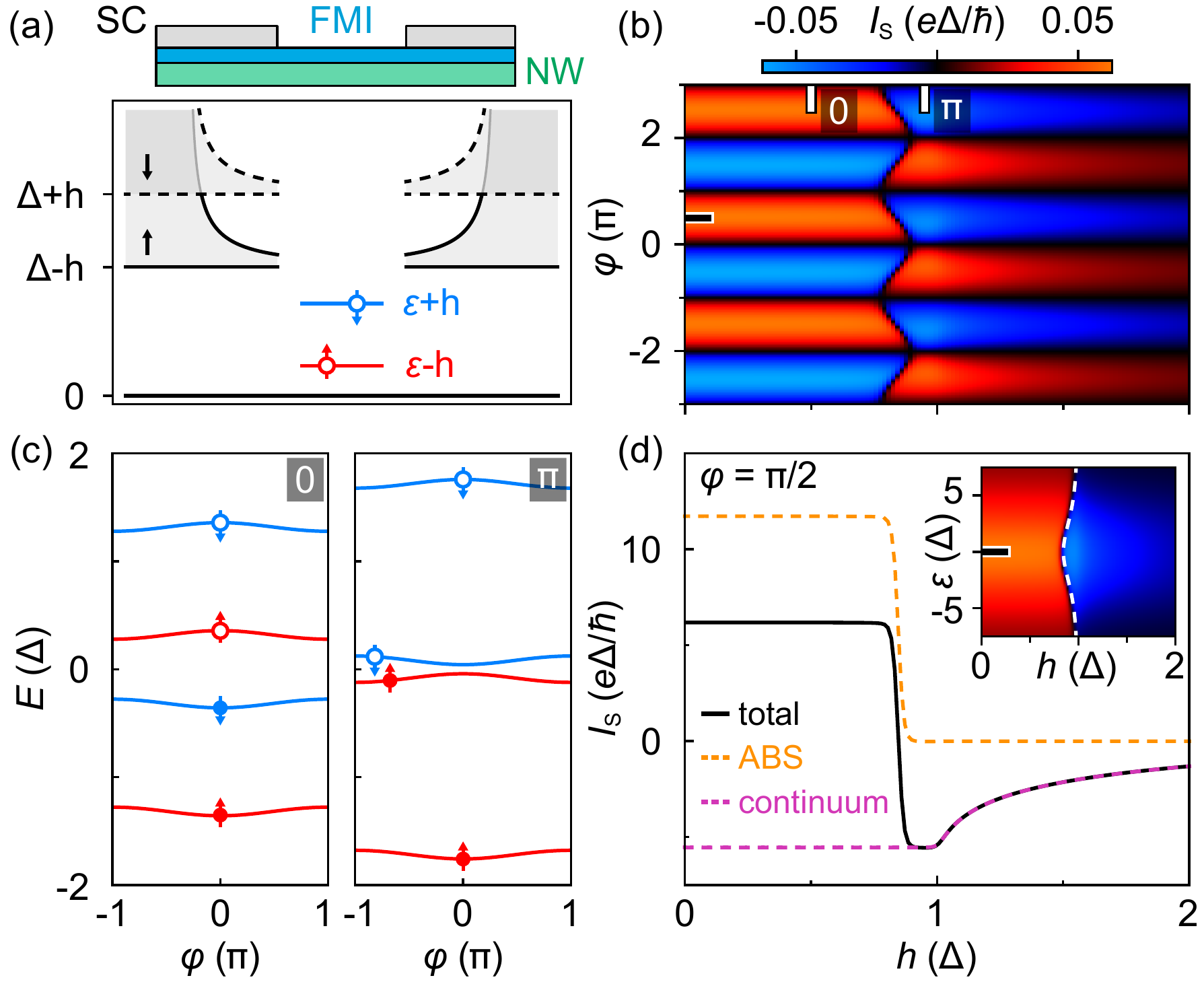}
\caption{\label{fig:4} 
(a) Schematic illustration of the modeled junction and the corresponding density of states with the induced superconducting gap in the wire leads, $\Delta$, and a discrete state in the junction at energy, $\varepsilon$, both spin-split by an exchange field, $h$.
(b)~Calculated supercurrent, $I_{\rm S}$, as a function of the superconducting phase difference across the junction, $\varphi$, and $h$, showing a 0\,--\,$\pi$ transition for $h\lesssim \Delta$.
(c) Energy of Andreev bound states (ABSs) in the junction as a function of $\varphi$ in the $0$ (left) and the $\pi$ (right) regimes.
(d) ABS and continuum contributions to the total $I_{\rm S}$ as a function of $h$ at $\varphi=\pi/2$ and $\varepsilon=0$.
The two contributions have opposite signs for $h\lesssim\Delta$, while the ABS contribution vanishes in the $\pi$ phase.
Inset: $I_{\rm S}$ dependence on $\varepsilon$, where the white line is an analytic expression for the 0\,--\,$\pi$ transition given in the Supplemental Material~\cite{Supplement}.
}
\end{figure}

These experimental observations suggest that the \mbox{0\,--\,$\pi$} transition is driven by a discrete flipping of the EuS domains affecting $\left<M\right>_\xi$ and changing the effective spin splitting of the ABSs in the junction.
We propose a simple transport model demonstrating that a sizable exchange field, arising from $\left<M\right>_\xi$, can polarize the junction, resulting in the $\pi$ phase shift.
The model describes a nanowire proximity-coupled to a ferromagnetic insulator and an interrupted superconductor, forming a Josephson junction [Fig.~\ref{fig:4}(a)].
The magnetic insulator induces an exchange field with amplitude $h(x)$, where $x$ is the position along the wire.
The superconductor induces a pairing potential, $\Delta(x)=\vert\Delta\vert e^{\pm i\varphi/2}$, in the lateral wire regions defining the leads, where $\varphi$ denotes the superconducting phase difference across the junction.
Assuming the short-junction limit, we describe the coupling between the leads by a single state, whose two spin-split levels are at energies $\varepsilon\pm h$. 
The extension to the long-junction limit including spin-orbit coupling yields qualitatively similar results and is discussed in Ref.~\cite{Maiani2022}.

We calculate the supercurrent through the junction, $I_{\rm S}$, proportional to $I_{\rm SW}$, using the Green functions formalism (see Supplemental Material~\cite{Supplement}).
In the tunneling regime, $I_{\rm S}$ displays a $\pi$ phase shift for $h\lesssim\Delta$ [Fig.~\ref{fig:4}(b)].
The supercurrent reversal can be understood by considering the field evolution of ABSs in the junction [Fig.~\ref{fig:4}(c)].
A finite $h$ lifts the spin degeneracy of the ABSs.
For large $h$, one ABS can cross zero energy, thus changing the junction ground state from anti-aligned to aligned spin configuration.
As a result, the ABS contribution to the supercurrent is suppressed, leaving only the subdominant transport via the continuum states with lower magnitude and opposite sign [Fig.~\ref{fig:4}(d)].
This can be understood by noting that the Josephson current is proportional to the product of the energy-dependent superconducting pairing in the leads $\sim\Delta^2/(\Delta^2-E^2)$, which becomes negative for energies $E>\Delta$~\cite{Rodero2011}.
In the experiment, the discrete domain flips lead to discontinuous jumps in the phase diagram. 

For low transmission, the field value of the 0\,--\,$\pi$ transition shows a relatively weak dependence on $\varphi$ and $\varepsilon$, see Figs.~\ref{fig:4}(b) and \ref{fig:4}(d) inset.
This is distinct from the interaction-driven supercurrent reversal in the quantum-dot regime~\cite{vanDam2006, Razmadze2020, Delagrange2016, Rozhkov2022}, where the transition can be smoothly tuned by an electrostatic gate changing the occupancy of the junction.
The quantum-dot scenario can be excluded since the $\pi$ phase shows no dependence on the junction-gate voltage (see Fig.~S5 in Supplemental Material~\cite{Supplement}).
Increasing the coupling rates to the leads results in a skewed CPR and a larger $h$ range where 0 and $\pi$ phases coexist (see Supplemental Material~\cite{Supplement}).

\begin{figure}[t!]
\includegraphics[width=\linewidth]{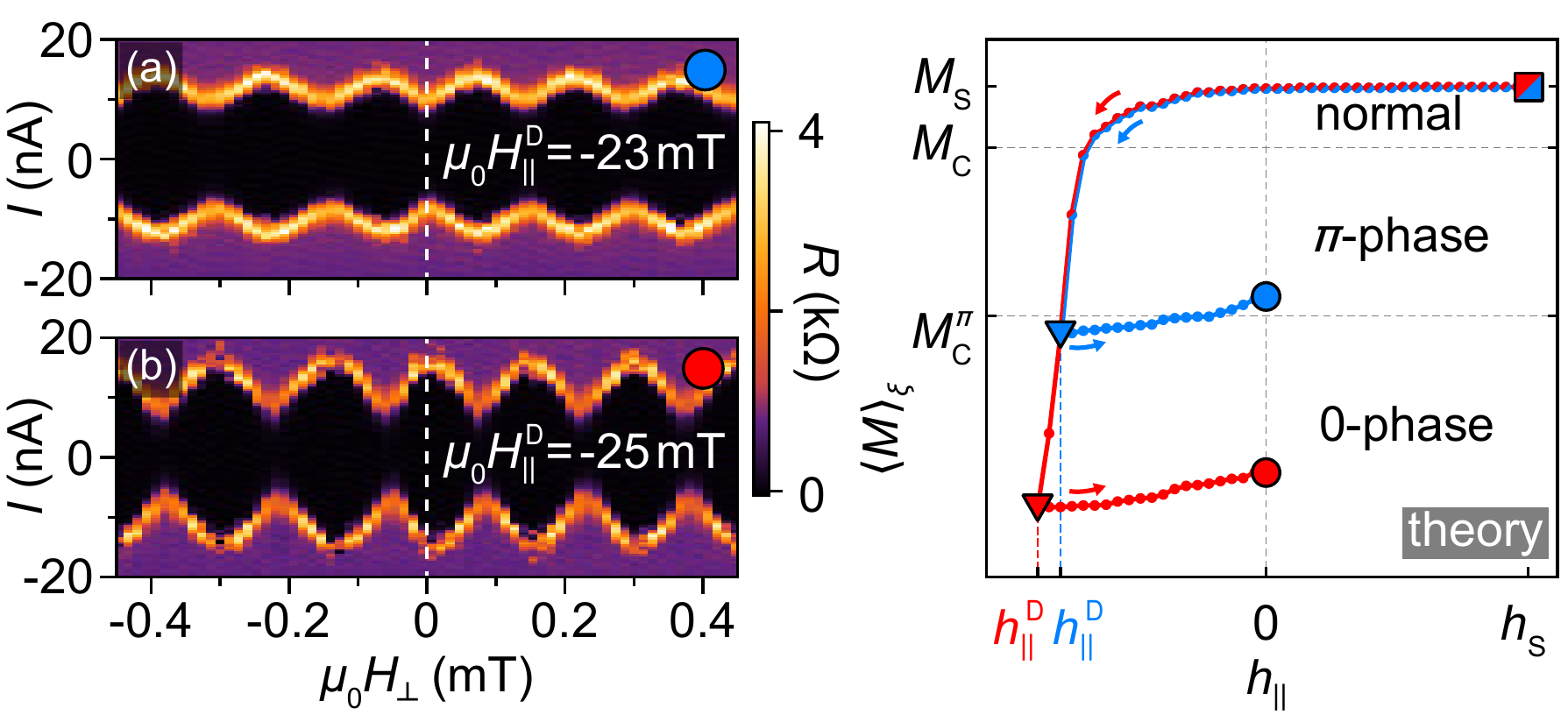}
\caption{\label{fig:5}
(a) Differential resistance, $R$, as a function of current bias, $I$, and flux-threading magnetic field, $H_\perp$, measured for $j_2^{\rm F}$\,-\,$j_3$ interferometer at zero parallel external field ($H_\parallel = 0$) showing a $\pi$-shifted current-phase relation.
The data were taken after polarizing the wire at $\mu_0 H_{\rm S}  = 100$~mT and demagnetizing it at $\mu_0 H_\parallel^{\rm D}=-23$~mT.
(b) Similar to (a) but taken after demagnetizing the wire at $\mu_0 H_\parallel^{\rm D}=-25$~mT, showing $0$-junction behavior.
(c) Calculated disorder-averaged induced magnetization, $\left< M \right>_\xi$, illustrating the experimental demagnetization scheme. After saturating $\left< M \right>_\xi$, the parameter representing the external magnetic field, $h_\parallel$, is swept to the variable demagnetization value, $h_\parallel^{\rm D}$, and then back to 0.
Depending on $h_\parallel^{\rm D}$, the junction can either relax to the $\pi$ or 0 phase.
}
\end{figure}

Finally, we demonstrate experimentally that the $\pi$ phase can be realized at zero external magnetic field by demagnetizing EuS with the following procedure [Fig.~\ref{fig:5}].
First, a saturating magnetic field, $\mu_0 H_{\rm S} = +100$~mT, was applied to fully polarize the EuS.
The field was then gradually swept through zero to a demagnetizing (negative) value, $H_\parallel^{\rm D}$, before returning to zero.
Carrying out the demagnetization loop for different $H_\parallel^{\rm D}$ values, we find that $j_2^{\rm F}$ transitions from $\pi$ to 0 phase as $\mu_0 H_\parallel^{\rm D}$ is changed from $-23$ to $-25$~mT [Figs.~\ref{fig:5}(a), \ref{fig:5}(b), and S8 in Supplemental Material~\cite{Supplement}].
This is similar to the behavior observed at the finite external field (Fig.~\ref{fig:3}), but now measured at $H_\parallel = 0$.
Qualitatively the same phenomenology was observed for $j_1^{\rm F}$ (see Fig.~S9 in Supplemental Material~\cite{Supplement}).
We ascribe this behavior to the $H_\parallel$-controlled EuS domain relaxation into a configuration with the remanent $\left<M\right>_\xi < M_{\rm C}$ as $H_\parallel$ is ramped back and forth.
The calculated demagnetization loops for two different demagnetization values support this picture [Fig.~\ref{fig:5}(c)].

In summary, we have studied the current-phase relation of triple-hybrid Josephson junctions comprised of semiconducting (InAs) nanowires with epitaxial ferromagnetic insulator (EuS) and superconductor (Al) shells.
The magnetic junctions showed a 0\,--\,$\pi$ phase transition within a hysteretic superconducting window in the parallel magnetic field.
We interpret the results in the context of magnetic domains and provide a simple theoretical model demonstrating that an induced average magnetization can account for the transition.
By demagnetizing the EuS layer, the $\pi$-phase can be realized at zero magnetic field, making the triple-hybrid junctions an attractive component for quantum and classical applications in superconducting circuitry.  

We thank Karsten Flensberg, Martin Leijnse, and Gorm Steffensen for valuable discussions, Claus S\o{}rensen for contributions to materials growth, and Shivendra~Upadhyay for assistance with nanofabrication. We acknowledge support from the Danish National Research Foundation, European Research Council (Grants Agreement No. 716655 and No. 856526), Microsoft, NanoLund, and a research grant (Project 43951) from VILLUM FONDEN.

\bibliography{bibfile}

\end{document}


\title{Supplemental Material: \\Supercurrent reversal in ferromagnetic hybrid nanowire Josephson junctions}
\author{D.~Razmadze}
\altaffiliation{These two authors contributed equally.}
\affiliation{Center for Quantum Devices, Niels Bohr Institute, University of Copenhagen, 2100 Copenhagen, Denmark}
\author{R.~Seoane~Souto}
\altaffiliation{These two authors contributed equally.}
\affiliation{Center for Quantum Devices, Niels Bohr Institute, University of Copenhagen, 2100 Copenhagen, Denmark}
\affiliation{Division of Solid State Physics and NanoLund, Lund University, 22100 Lund, Sweden}
\author{L.~Galletti}
\affiliation{Center for Quantum Devices, Niels Bohr Institute, University of Copenhagen, 2100 Copenhagen, Denmark}
\author{A.~Maiani}
\affiliation{Center for Quantum Devices, Niels Bohr Institute, University of Copenhagen, 2100 Copenhagen, Denmark}
\author{Y.~Liu}
\affiliation{Center for Quantum Devices, Niels Bohr Institute, University of Copenhagen, 2100 Copenhagen, Denmark}
\author{P.~Krogstrup}
\affiliation{Center for Quantum Devices, Niels Bohr Institute, University of Copenhagen, 2100 Copenhagen, Denmark}
\author{C.~Schrade}
\affiliation{Center for Quantum Devices, Niels Bohr Institute, University of Copenhagen, 2100 Copenhagen, Denmark}
\author{A.~Gyenis}
\altaffiliation{Current address: Department of Electrical, Computer \& Energy Engineering, University of Colorado Boulder, CO 80309, USA}
\affiliation{Center for Quantum Devices, Niels Bohr Institute, University of Copenhagen, 2100 Copenhagen, Denmark}
\author{C.~M.~Marcus}
\affiliation{Center for Quantum Devices, Niels Bohr Institute, University of Copenhagen, 2100 Copenhagen, Denmark}
\author{S.~Vaitiek\.{e}nas}
\affiliation{Center for Quantum Devices, Niels Bohr Institute, University of Copenhagen, 2100 Copenhagen, Denmark}

\date{\today}

\maketitle

\setcounter{figure}{0}
\setcounter{equation}{0}
\renewcommand\thefigure{S\arabic{figure}}
\renewcommand{\tablename}{Table.~S}
\renewcommand{\thetable}{\arabic{table}}
\twocolumngrid

\section{Sample preparation}
Both wires A and B had hexagonal InAs cores that were grown to a length of roughly $10~\mu$m and a diameter of 120~nm using molecular beam epitaxy~\cite{Krogstrup2015}.
For wire A, a two-faceted EuS (4~nm) shell was grown in a separate metal-deposition chamber without breaking the vacuum; for both batches, three-faceted \textit{in~situ} Al (10~nm for wire A and 7~nm for wire B on the middle facet) shells were grown in the III-V growth chamber~\cite{Liu2019,Liu2019_2}.
Devices were fabricated on a Si chip with 200~nm SiOx capping.
The junctions were formed by selectively wet etching short segments of Al ($\sim100$~nm), leaving EuS intact.
After metallization of \textit{ex~situ} Al (190~nm) ohmic contacts, the wires were coated with HfO$_2$ (6~nm) dielectric layer, followed by the deposition of Ti/Au (5/195~nm) top-gate junction electrodes for the electrical control of each junction.
The main multi-interferometer device is shown in the main-text Fig.~1(a).
The second multi-interferometer was lithographically similar to the main device, but with only one reference (non-magnetic) junction.
More detailed description of wire growth and device fabrication is given in Ref.~\cite{Vaitiekenas2020}.

\section{Measurements}
Four-terminal differential resistance measurements were performed using standard lock-in techniques with an ac current excitation of 0.5~nA at $\sim33$~Hz, in a dilution refrigerator with a three-axis (1, 1, 6)-T vector magnet and base temperature of 20~mK.
In total 14 multi-interferometer devices with nanowires from the wire A growth batch were investigated: 10 devices were either not electrically connected or did not superconduct, 2 devices did not show flux dependence, and 2 devices worked as intended and are reported in this work.
The $I-V$ characteristic curves for these two devices, taken at zero-field cooling and before applying the gate voltages or external magnetic field, are shown in Fig.~\ref{fig:S10}.

\section{Magnetization anisotropy}

Because of the highly anisotropic shape of the measured wires, we expect the EuS to have an easy magnetization axis along the wire, as was observed on similar nanowires~\cite{Vaitiekenas2020}.
We verify this by measuring differential resistance, $R$, for $j_2^{\rm F}$ of the main multi-interferometer device as a function of current bias, $I$, and external magnetic field, $H_\theta$, applied at three different in-plane angles, $\theta$ = 0, 30, and 60 degrees with respect to wire A axis [Fig.~\ref{fig:S11}(a)-(c)].
We find that the magnitude of the coercive field, $H_{\theta, \rm C}$, taken at the field value where the superconducting features are the most pronounced, increases as 1/cos$(\theta)$, see Fig.~\ref{fig:S11}(d).
This suggests that it is predominantly the parallel component of the applied field that controls the magnetization.
We note that $R$ increases within the superconducting window for $\theta = 60$~degree.
Such an increase in \textit{differential} resistance can be interpreted as a measurement artifact in the case of the supercurrent, suppressed by a finite off-axis (in-plane) field component, decreasing to a value comparable to the ac current excitation.
The superconducting phase uncertainty, discussed in the main text, enhances this effect.

\section{Wire parameters}\label{Sec:wireParam}

To characterize the magnetic and superconducting properties of the studied wires, we perform further characterization measurements on two additional devices. 
We begin by investigating the superconducting properties of the coupled Al and EuS shells. 
Differential shell resistance, $R_{\rm S}$, measured in a four-terminal configuration for the shell-characterization device [Fig.~\ref{fig:S12}(a)] as a function of external magnetic field, $H_{\parallel}$, and temperature, $T$, shows a skewed superconducting phase diagram with a characteristic critical temperature $T_{\rm C0}=(1.2\pm0.1)$~K around the coercive field [Fig.~\ref{fig:S12}(b)].
The shell
resistivity is given by $\rho_{\rm Al} = R_{\rm N} A_{\rm Al}/ L = (80\pm10)~\Omega$\,nm, where $R_{\rm N} = (150\pm2)$~$\Omega$, is the normal state resistance [Fig.~\ref{fig:S12}(c)], $A_{\rm Al} = (1200\pm200)$~nm$^2$ is the Al shell cross-section, and $L=(2200\pm100)$~nm is the distance between the voltage probes.
The Drude mean free path for electrons in the Al shell is determined using $l_{\rm e} = m_{\rm e}{ }v_{\rm F}/ e^2 n \rho_{\rm Al} = (5\pm1)$~nm, with electron mass $m_{\rm e}$, electron Fermi velocity in Al $v_{\rm F} = 2.03 \times 10^6$~m/s \cite{Kittel2005}, electron charge $e$ and charge carrier density $n = k_{\rm F}^3/ 3 \pi^2$, where $k_{\rm F}$ is the Fermi wave vector.
With this, the dirty-limit superconducting coherence length in Al is given by~\cite{Tinkham1996, Vaitiekenas2019} $\xi~=~\sqrt{\pi \hbar v_{\rm F}  l_{\rm e}/24 k_{\rm B} T_{\rm C0}} = (90\pm10)$~nm, where $\hbar$ is the reduced Planck constant and $k_{\rm B}$ is the Boltzmann constant.

Next, we investigate the magnetic and superconducting properties proximity-induced into InAs using the tunneling-spectroscopy device [Fig.~\ref{fig:S12}(d)].
Differential tunneling conductance, $G_{\rm T}$, measured across the gate-controlled tunnel barrier at the end of the hybrid nanowire into a normal contact as a function of $H_\parallel$ and source-drain bias, $V_{\rm T}$, shows a characteristic superconducting window away from $H_\parallel = 0$ [Fig.~\ref{fig:S12}(e)], consistent with the current-bias measurements presented in the main-text Fig.~1.
Within the narrow window, we observe a superconducting gap, maximal at $\mu_0 H_\parallel = -24$~mT, with two coherence peaks around $\pm 50$ and $\pm80~\mu$eV [Fig.~\ref{fig:S12}(f)].
We interpret these features to be the induced superconducting gap, $\Delta \approx 65~\mu$eV, spin-split by an effective exchange field, $h_0 \approx 15~\mu$eV.
We note that the lower-energy coherence peaks are larger, which could be explained by strong spin-orbit coupling~\cite{Bruno1973}.

\section{Domain size estimation}
The magnetic properties induced in the superconductor by the proximity of a ferromagnetic insulator decay exponentially on a length scale dictated by $\xi$ away from the interface~\cite{Tokuyasu1988}.
In other words, the exchange field at a given point in the superconductor is proportional to the local magnetization in the ferromagnetic insulator weighted by an exponential distribution.
Because the thickness of the superconductor is much smaller than $\xi$, we consider the effective exchange field in the superconductor $h$, proportional to the average magnetization $\left< M \right>_\xi$, to be homogeneous along the direction normal to the interface and to vary only along the wire (x~axis) as
\begin{equation}
    h(x) = \zeta_{\rm h}(x)\,*\,e^{-|x|/\xi} = \int_{-\infty}^{\infty} \zeta_{\rm h}(x') e^{|x - x'|/\xi}\,\mathrm{d}x'\,,
    \label{eq:indeuced_zeeman}
\end{equation}
where $\zeta_{\rm h}$ is the exchange field induced locally by an infinitesimally short segment of ferromagnetic insulator, and $\,*$ denotes convolution.
For simplicity, we consider domains of minimal constant size, $d$, with magnetization pointing along the wire, that can be flipped by an external magnetic field applied anti-parallel to the  magnetization.

For external fields well above the coercive field, where the local magnetization is saturated and all the domains are aligned (Fig.~\ref{theory_S1}), $\zeta_{\rm h}(x)$ is constant, $\zeta_{\rm S}$, and so the saturation exchange field is given by
\begin{equation}
    h_{\rm S}=\zeta_{\rm S}\int_{-\infty}^{\infty}\mathrm{d}x'\, e^{-|x'|/\xi}\to \zeta_{\rm S}=\frac{h_{\rm S}}{2\xi}\,.
    \label{magnetization_homogeneous}
\end{equation}
Experimentally, we find that the superconductivity is suppressed in the fully-polarized configuration. This implies that the average magnetization exceeds the Chandrasekhar-Clogston limit \cite{Chandrasekhar1962,Clogston1962}, putting a bound on $h_{\rm S}\geq\Delta/\sqrt{2}$.

At the coercive field, the domain configuration is maximally randomized, which can be represented by the antiparallel alignment of magnetization in neighboring domains (Fig.~\ref{theory_S1}), resulting in a minimal average magnetization.
In this case, $\zeta_{\rm h}(x)=\pm\zeta_{\rm S}$ depending on the local magnetization orientation.
At the center of a domain, taken to be \mbox{$x=0$}, Eq.~\eqref{eq:indeuced_zeeman} becomes
\begin{equation}
    h(x=0)=2\,\xi\,\zeta_{\rm S}\left(1-2\frac{e^{-d/2\xi}}{1+e^{d/\xi}}\right)\,.
    \label{Magnetization_1}
\end{equation}
The average effective exchange field of the sample at the coercive field can be calculated by integrating absolute value of Eq. \eqref{eq:indeuced_zeeman} over a domain, giving
\begin{equation}
    \begin{split}
    h_0 &= \frac{1}{d}\int_{-d/2}^{d/2} \left|h(x)\right| \,\mathrm{d}x =\\
    &= h_{\rm S}\left[1-\frac{4\xi}{d}\frac{e^{-d/2\xi}}{1+e^{d/\xi}}{\rm sinh}(d/2\xi)\right]\,,
\label{Magnetization_2}
    \end{split}
\end{equation}
which, together with the bound for $h_{\mathrm S}$ and the values for $\Delta$ and $h_0$ deduced in the previous section, yields \mbox{$d\lesssim 0.4\,\xi$}.
Using the estimated $\xi$ in Al, we get $d\lesssim 35 $~nm, but note that the induced superconducting coherence length might be longer.

The above estimate for the upper domain-size limit is based on a conservative bound on the saturation magnetization $h_{\rm S}>\Delta/\sqrt{2}$.
As the superconducting window in the experiment appears away from $H_\parallel = 0$, we expect an even larger $h_{\rm 0}$, and so a smaller $d$.
In addition, we expect that the parent Al gap that sets the  Chandrasekhar-Clogston limit is larger than the induced gap, which would further increase the limit for $h_0$.
Finally, any size variation of the domains (which is a more realistic scenario) would increase $h_0$ and decrease $d$.
Therefore, we conclude that the measured wires are in the regime $d<\xi$.

\section{Magnetic-Domain Model}

To qualitatively explain the observed hysteretic superconductivity, we employ a kinetic random-field Ising model~\cite{Sethna1993}.
We model a magnetic insulator region of size $\xi^2$ by a microscopic grid of $10 \times 10$ magnetic grains (consistent with the estimate for the upper domain-size limit) with periodic boundary conditions.
The average of the grid magnetization represents $\left< M \right>_\xi$.
The system is described by
\begin{equation}
    H = -\frac{1}{2}\sum_{ij} J_{ij} S_i S_j - \sum_i \left(h_\parallel S_i + f_i S_i\right) \,,
    \label{eq:krfim_hamiltonian}
\end{equation}
where $S_i = \pm 1$ is the Ising spin, $J_{ij}$ is the exchange coupling that we assume to be 1 for the nearest neighbors and zero otherwise, $h_\parallel$ is the external magnetic field, and $f_i \sim \mathcal{N}(\mu = 0, \sigma^2 = 1)$ is a random field that accounts for the magnetic-domain disorder.
To take dissipative dynamics into account, we use Glauber algorithm~\cite{Glauber1963}, which, after $h_\parallel$ is changed, iteratively flips randomly-chosen spins with probability
\begin{equation}
    p = \frac{1}{2}\left[1 - \tanh \left(\frac{\Delta E}{2 T}\right)\right]\,.
    \label{eq:glauber_dynamics}
\end{equation}
Here, $\Delta E$ is the change in energy of the system due to the spin-flip and $T$ is the temperature.
For simplicity, we consider $T=0$, in which case $p$ is either 0 or 1, depending on the sign of $\Delta E$.

We initialize the grid in a fully spin-polarized configuration ($S_i = -1$) and sweep $h_\parallel$ from -3 to 3 and back, allowing the system to evolve for 4000 iterations at each step.
The results of 20 different $f_i$ realizations are shown in Fig.~\ref{theory_S2}(a).
In a typical scenario, a few domains flip after $h_\parallel$ passes through 0, causing small changes in $\left< M \right>_\xi$, which is followed by an abrupt domain-flip avalanche resulting in a fully polarized grid, see Fig.~\ref{theory_S2}(b).
The disorder-averaged hysteresis curve shown in Fig.~\ref{theory_S2}(c) further illustrates how $\left< M \right>_\xi$ first decreases smoothly as $h_\parallel$ is changed, but then jumps through 0 and inverts abruptly around the coercive field.
This asymmetry can be further illustrated by considering the probability, $p$, for the grid to have a particular $\left< M \right>_\xi$ value.
Figure~\ref{theory_S2}(d) shows a histogram of $\left< M \right>_\xi$ take from the 20 down-sweeps in Fig.~\ref{theory_S2}(a).
To put these results into the context of our experiment, we introduce two critical-magnetization values: $\left|M_{\rm C}\right|=0.9\left< M \right>_\xi$, below which the system is superconducting, and $\left|M_{\rm C}^\pi\right|=0.7\left< M \right>_\xi$, above which the system is in the $\pi$ phase.
Based on these findings, it is apparent that the probability of the $\pi$-phase formation is higher at the onset compared to the end of the superconducting window.

The asymmetry of the hysteresis curve around the coercive field is presumably the reason for the observed $\pi$-phase at the onset of the superconducting window, but not at the end.

\section{Tansport Model}
A minimal model describing the junction and used in the main text is given by the Hamiltonian
\begin{equation}
    H=H_{\rm L}+H_{0}+H_{\rm T}\,.
\end{equation}
The superconducting parts of the nanowire, that we refer to as leads, are described by
\begin{equation}
    H_{\rm L}=\sum_{\nu,k}\hat{\Psi}^{\dagger}_{\nu k}\hat{\mathcal{H}}_{\nu k}\hat{\Psi}_{\nu k}\,,
    \label{H_leads}
\end{equation}
where $\hat{\Psi}^{\dagger}_{\nu k}=(c_{\nu k\uparrow}^{\dagger},c_{\nu k\downarrow})$ is the Nambu spinor, with the electron annihilation operator $c$, the lead index \mbox{$\nu \in \{L, R\}$}, the electron quasi-momentum~$k$, and the charge carrier spin $\uparrow, \downarrow$.
In Eq. \eqref{H_leads}, $\hat{\mathcal{H}}_{\nu k}=\epsilon_{\nu k}\hat{\tau}_z+h_{\nu}\hat{\tau}_0+\Delta_\nu\hat{\tau}_x$, where $\epsilon$ is the electron energy, $h$ is the spin-splitting field, and $\Delta$ is the pairing potential.
The Pauli matrices $\hat{\tau}_{x,y,z}$ act in the Nambu space.
Assuming the short-junction limit, we describe the normal part between the two superconductors by a single normal level,
\begin{equation}
    H_{0}=\hat{\Psi}^{\dagger}_{0}\hat{\mathcal{H}}_{0}\hat{\Psi}_{0}\,,
\end{equation}
where $\hat{\mathcal{H}}_0=\varepsilon\hat{\tau}_z+h_{\rm J}\hat{\tau}_0$, with the energy of the level $\varepsilon$ and the spin-splitting field in the junction $h_{\rm J}$.
Lastly, the tunneling between the two superconductors is given by
\begin{equation}
    H_{\rm T}=\sum_{\nu,k}\hat{\Psi}^{\dagger}_{\nu k}\hat{V}_{\nu k}\hat{\Psi}_{0}+\mbox{H.c.}\,,
\end{equation}
where $\hat{V}_{\nu k}=V_{\nu k}\hat{\tau}_z e^{i\hat{\tau}_z\varphi_\nu}$ is the tunneling matrix, with the superconducting phase $\varphi_\nu$ in lead $\nu$ and the tunneling amplitude~$V_\nu$. We define the superconducting phase difference as $\varphi=\varphi_L-\varphi_R$.
We consider $V_\nu$ to be momentum-independent, in which case the tunneling rate to the lead~$\nu$ is given by $\Gamma_\nu=\pi|V_\nu|^2\rho_\nu$, where $\rho_\nu$ is the normal density of states in the leads at the Fermi level, and $\Gamma=\Gamma_{\rm L}+\Gamma_{\rm R}$.
Hereafter, we use $e=\hbar=k_{\rm B}=1$.

\section{Green function formalism}
To describe the transport through the system, we use the Green function formalism, summarized in Ref. \cite{Rodero2011}.
The retarded/advanced (r/a) Green function of the central (junction) region is given by
\begin{equation}
    \hat{G}^{r/a}_0(\omega)=\left[\hat{g}_{0}^{-1}(\omega)-\hat{\Sigma}^{r/a}_L(\omega)-\hat{\Sigma}^{r/a}_R(\omega)\right]^{-1},
    \label{Eq:Dyson}
\end{equation}
where $\hat{g}_{0}^{-1}(\omega)=(\omega+h_{\rm J})\hat{\tau}_0+\varepsilon\hat{\tau}_z$ describes the isolated normal region as a function of the electron energy $\omega$, and $\hat{\Sigma}^{r/a}_\nu$ is the self-energy describing the coupling to the lead $\nu$, given by
\begin{equation}
    \hat{\Sigma}^{r/a}_{\nu}(\omega)=\sum_\nu\Gamma_\nu\left[\mathfrak{g}^{r/a}_{\nu}(\omega)\hat{\tau}_0+\mathfrak{f}^{r/a}_{\nu}(\omega)\tau_x e^{\hat{\tau}_y\varphi_\nu}\right]\,.
\end{equation}
Here 
\begin{equation}
\begin{split}
    \mathfrak{g}^{r/a}_{\nu}(\omega)&=-\frac{\omega+h_{\nu}\pm i\eta}{\sqrt{\Delta^2-(\omega+h_{\nu}\pm i\eta)^2}}\,,\\
    \mathfrak{f}^{r/a}_{\nu}(\omega)&=\frac{\Delta}{\sqrt{\Delta^2-(\omega+h_{\nu}\pm i\eta)^2}}\,,
    \label{Eq::GFs}
\end{split}
\end{equation}
and $\eta$ is the Dynes parameter, controlling the width of the superconducting coherent peaks at $\omega=\pm\Delta$.

The non-equilibrium Green function is given by 
\begin{eqnarray}
    \hat{G}^{+-}_{0}(\omega)= \left[G^a(\omega)-G^r(\omega)\right]n_F(\omega)\,,
\end{eqnarray}
where $n_F$ is the Fermi distribution function.
With this, we can calculate the spin polarization of the junction, $S_{\rm Z} = (n_{\uparrow} - n_{\downarrow})$, where the spin-up and spin-down populations are given by
\begin{equation}
\begin{split}
    n_{\uparrow}&=-\mbox{Im}\left\{\frac{\mbox{1}}{2\pi}\int d\omega \left[\hat{G}^{+-}_{0}(\omega)\right]_{11}\right\}\,,\\
    n_{\downarrow}&=1+\mbox{Im}\left\{\frac{\mbox{1}}{2\pi}\int d\omega \left[\hat{G}^{+-}_{0}(\omega)\right]_{22}\right\}\,,
\end{split}
\end{equation}
with the subindex $1,2$ referring to the Nambu component. Finally, the supercurrent at $\nu$ interface of the junction is given by
\begin{equation}
\begin{split}
    I_\nu= \mbox{Re}
    \left\{ \int \right. d\omega\, \frac{n_F(\omega)}{2\pi}\mbox{Tr} & \left[ \hat{\Sigma}_\nu(\omega)\cdot\hat{G}^{+-}_{0}(\omega)  \right. \\
    & \left. -\hat{G}^{+-}_{0}(\omega)\cdot\hat{\Sigma}_\nu(\omega)  \right] \left.\vphantom{\int} \right\} \,,
\end{split}
\end{equation}
where the trace is taken over the Nambu space and \mbox{$I_{\rm L}=-I_{\rm R} \equiv I_{\rm S}$}.

In the experiment, the measured $I_{\rm SW}$ is usually less than $I_{\rm S}$~\cite{Tinkham1996}.
The reduction is due to stochastic jumps between the superconducting and resistive branches when the bias current approaches $I_{\rm S}$.
In the absence of thermally activated processes $I_{\rm SW} \approx I_{\rm S}$.
Increasing  thermal fluctuations leads to a spread of $I_{\rm SW}$ values whose mean is proportional to $I_{\rm S}$.
In our case, $k_{\rm B} T \sim 2~\mu$eV (considering the fridge base temperature $T=20$~mK) is a few times smaller than the Josephson energy $E_{\rm J} = \hbar I_{\rm SW}/2e \sim 10~\mu$eV of a junction with $I_{\rm SW} = 5$~nA [see Fig.~3(a) in the main text].

\section{Analytic expression}
The position of the ABSs is given by the condition $\mbox{det}[\hat{G}_0(\omega)]=0$~[see Eq.~\eqref{Eq:Dyson}].
The transition between 0 and $\pi$ phases takes place when one of the spin-split ABSs crosses zero energy, that is when
\begin{equation}
    \mbox{det}[\hat{G}^{-1}_0(\omega=0)]=0\,.
    \label{eq:det}
\end{equation}
Solving Eq. \eqref{eq:det} for $h_{\rm L}=h_{\rm R}=h_{\rm J}\equiv h$ we find that the 0\,--\,$\pi$ transition occurs when
\begin{equation}
    \varepsilon=\sqrt{\left[h-\Gamma\, \mathfrak{g}^r(0)\right]^2-\left[b\,\mathfrak{f}^r(0)\right]^2}\,,
    \label{Eq:solDet}
\end{equation}
with $h\leq\Delta$ and
\begin{equation}
    b=\left|\sum_\nu \Gamma_\nu e^{i\varphi_\nu}\right|\,.
\end{equation}
The solution becomes especially simple for $h_{\rm L}=h_{\rm R}=0$, where $\mathfrak{g}^r(0)=0$ and $\mathfrak{f}^r(0)=1$, in which case Eq. \eqref{Eq:solDet} can be written as
\begin{equation}
    \varepsilon=\sqrt{h_{\rm J}^2-b^2}\,.
\end{equation}

\section{Supporting theory results}
In the following, we provide additional theory results of our study.
We begin by analyzing $I_{\rm S}$ in the single-barrier regime ($\Gamma_{\rm L} \gg \Gamma_{\rm R}$) as a function of $h$ and $\varepsilon$~(Fig.~\ref{theory_S3}).
The total $I_{\rm S}$ shows a 0\,--\,$\pi$ transition for $h\lesssim\Delta$, well described by Eq.~\eqref{Eq:solDet} [see the white dashed line in Fig.~\ref{theory_S3}(a)];
the dependence on $\varepsilon$ is weak due to the small channel transmission.
The total $I_{\rm S}$ consists of two oppositely directed currents carried by the ABSs~[Fig.~\ref{theory_S3}(b)] and the continuum of states~[Fig.~\ref{theory_S3}(c)].
The current contribution from ABSs dominates for small $h$ values, but it vanishes in the $\pi$ phase, as a consequence of the parity change in the junction.
The current contribution from the continuum of states with  an opposite sign dominates the transport in the $\pi$ phase and, in our model, is the main cause of the $I_{\rm S}$ reversal.

In case of perfect transmission ($\Gamma_{\rm L} = \Gamma_{\rm R}$), the 0\,--\,$\pi$ transition in $h$ depends strongly on $\varphi$ [Fig.~\ref{theory_S4}(a)].
In the open regime ($\Gamma\gg\Delta$), the system transitions between the two ground states as a function of $\varphi$ for $0<h<\Delta$.
For $h\sim 0$, the $0$ phase dominates, whereas for $h\sim\Delta$ the system is predominantly in the $\pi$ phase.
At the intermediate $h$ values, the system is in the $0'$ or $\pi'$ phases, depending on the global energy minimum~\cite{Rodero2011}.
We note that for $\varepsilon\gg\Delta,\,\Gamma$ the transition approaches $h=\Delta$~[Fig.~\ref{theory_S4}(b)], similar to the single-barrier regime~(Fig.~\ref{theory_S3}).
This is due to the reduction of the junction transmission with $\varepsilon$.

For completeness, we also investigate the quantum dot (QD) regime ($\Gamma\ll \Delta$), see Fig.~\ref{theory_S5}.
We consider a non-interacting quantum dot with a single spin-split energy level weakly coupled to the spin-split superconducting leads.
For $\varepsilon<\Delta$, the 0\,--\,$\pi$ transition takes place at $h\sim\varepsilon$, when the spin-split level crosses zero energy, whereas for $\varepsilon>\Delta$ the transition saturates at $h\lesssim\Delta$, as the transmission is suppressed.
In this limit, the ABS contribution (in the 0-phase) is maximal at $\varepsilon=0$, while the one from the continuum of states contribution peaks for $\varepsilon\approx\pm\Delta$.
This is because of the different resonant conditions for the two contributions to the total current.

Usually, the 0\,--\,$\pi$ transition is associated with a parity change in the junction, where the ground state transits from spin singlet to doublet.
This is to be contrasted with our model, where the transition is due to the spin polarization of the ground state, induced by the exchange field.
The boundary between the spin-degenerate ($S_{\rm Z} = 0$) and spin-polarized ($S_{\rm Z} > 0$) junction agrees well with the \mbox{0\,--\,$\pi$} transition [Eq.~\eqref{Eq:solDet}] for $h<\Delta$ in all three (quantum dot, single barrier, and open) regimes~[Fig.~\ref{theory_S6}].
For $h>\Delta$, the spin-splitting of the induced gap can reverse $I_{\rm S}$, without polarizing the junction.
This is best illustrated for $\vert\varepsilon\vert>h$ in the QD regime [Fig.~\ref{theory_S6}(a)].

So far, we have considered a homogeneous spin-splitting across the system ($h_{\rm L} = h_{\rm R} = h_{\rm J} \equiv h$).
Because of the interruption in the superconductor, it is possible that the average magnetization in the junction, where the Al is removed, is different from the one in the leads~\cite{Liu2021}.
Figure~\ref{theory_S7} shows 0\,--\,$\pi$ phase diagrams for the three considered regimes, as a function of the spin-splitting field in the junction, $h_{\rm J}$, and the leads, $h_{\rm L, R}$.
In the QD regime, a small $h_{\rm J}$ can reverse the $I_{\rm S}$, but a large $h_{\rm L, R} \sim \Delta$ is needed for the $\pi$ phase to appear if $h_{\rm J}=0$~[Fig.~\ref{theory_S7}(a)].
In general, the onset of the $\pi$ phase appears at lower field values for parallel $h_{\rm J}$ and $h_{\rm L, R}$ compared to the anti-parallel configuration.
This is more apparent in the single-barrier and the open regimes~[Figs. \ref{theory_S7}(b) and \ref{theory_S7}(c)], where the transition is rather insensitive to $h_{\rm J}$, taking place for $h_{\rm L, R}\sim \Delta$.

Finally, we show that the softening of the induced gap, for example, due to poor proximity or spin-flip scattering, does not change the qualitative features of the 0\,--\,$\pi$ transition.
In our model, the softening of the gap is introduced through the Dynes parameter [$\eta$ in Eqs. \eqref{Eq::GFs}], which smoothens the superconducting coherence peaks at $\pm(\Delta\pm h)$, leading to a finite subgap density of states [Figs.~\ref{theory_S8}(a) and \ref{theory_S8}(b)].
A finite $\eta$ blurs out the boundary between the 0 and $\pi$ phases, shifting the transition to slightly larger $h$ values.

\begin{figure*}[t!]
\includegraphics[width=\linewidth]{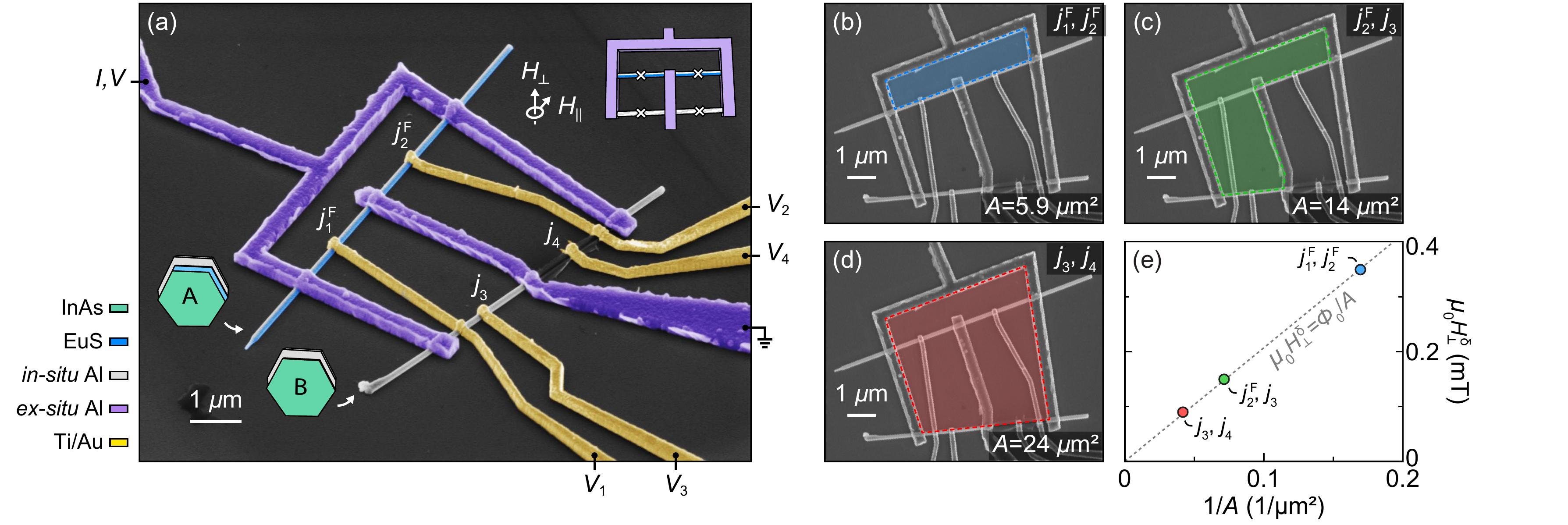}
\caption{\label{fig:S1}
(a) Color-enhanced, tilted-view micrograph of the main multi-interferometer device comprised of ferromagnetic (target) and non-magnetic (reference) nanowires.
The insets show schematic device layout and wire cross-sections.
The segment of wire~B with $j_4$ was damaged during the measurements due to electrostatic discharge.
\mbox{(b)-(d)} Top-view micrographs of the main device with highlighted effective loop areas, $A$, used for the current-phase relation measurements presented in the main-text Fig.~2.
The different loops were formed by opening (b) two ferromagnetic, (c) one ferromagnetic and one non-magnetic, and (d) two non-magnetic junctions, while keeping the other two junctions closed. 
(e) Switching current periods, $\mu_0 H_\perp^\delta$, deduced for different loops from the data shown in the main-text Fig.~2, grows as $\Phi_0/A$, where $\Phi_0 = h/2e$ is the superconducting flux quantum and $A$
is measured from the micrographs.}
\end{figure*}

\begin{figure*}[t!]
\includegraphics[width=\linewidth]{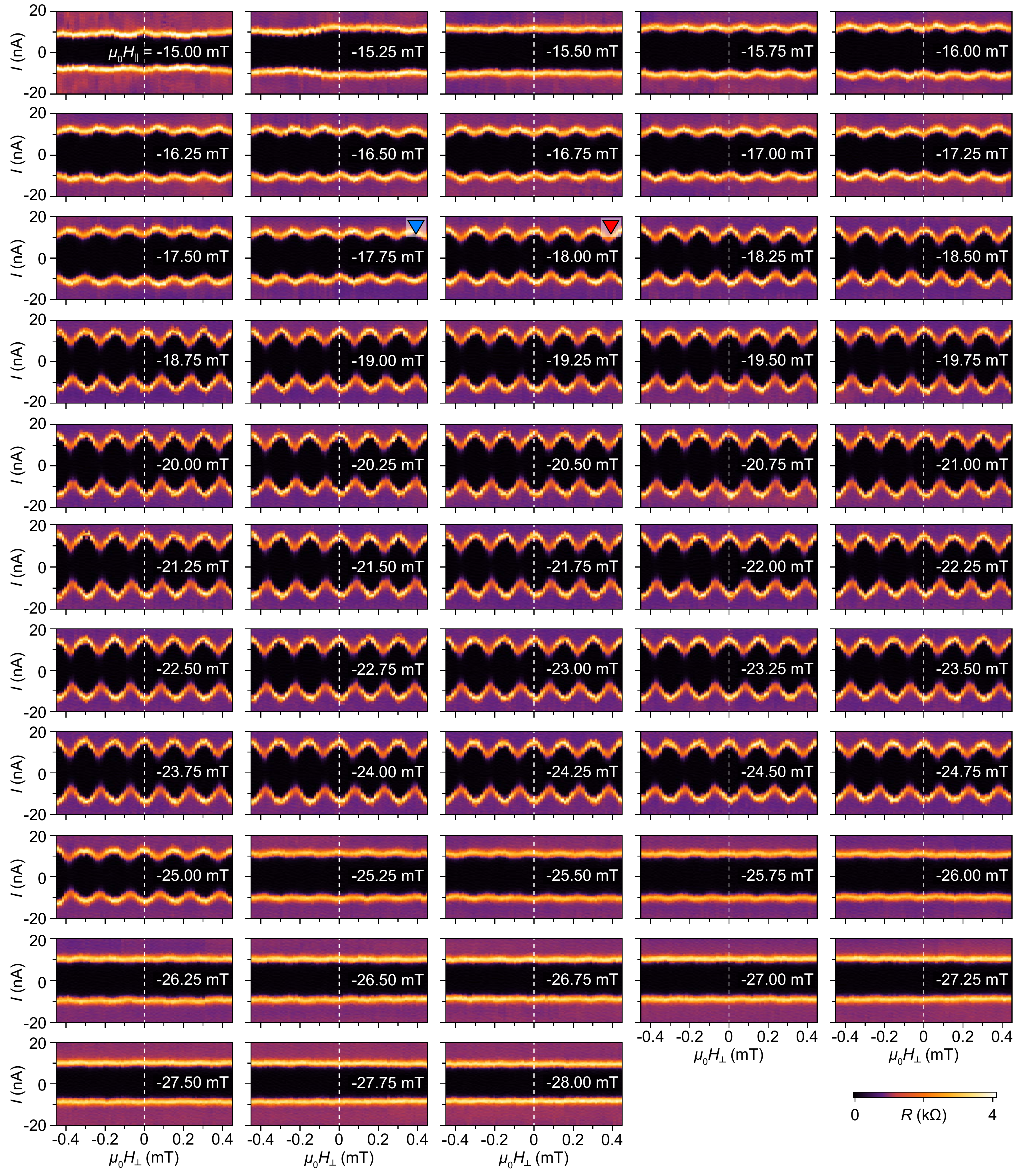}
\caption{\label{fig:S2}
Differential resistance, $R$, of the $j_2^{\rm F}$\,-\,$j_3$ interferometer as a function of current bias, $I$, and flux-threading perpendicular magnetic field, $H_\perp$, measured at decreasing parallel external magnetic field, $H_\parallel$, after polarizing the wire at $\mu_0 H_\parallel = 100$~mT.
The interferometer displays a periodic switching current oscillations in the range between $\mu_0 H_\parallel \sim -15.5$ and $-26.5$~mT, with an abrupt shift of the relative phase from $\pi$ to 0 around $\mu_0 H_\parallel = -18$~mT.
The data were taken at junction-gate voltages $V_1 = -6$~V, $V_2 = 5$~V, $V_3 = 0.5$~V, and $V_4 = 0$ (after the electrostatic discharge damage of $j_4$).
}
\end{figure*}

\begin{figure*}
\centering
\begin{minipage}[t]{.47\textwidth}
\includegraphics[width=\linewidth]{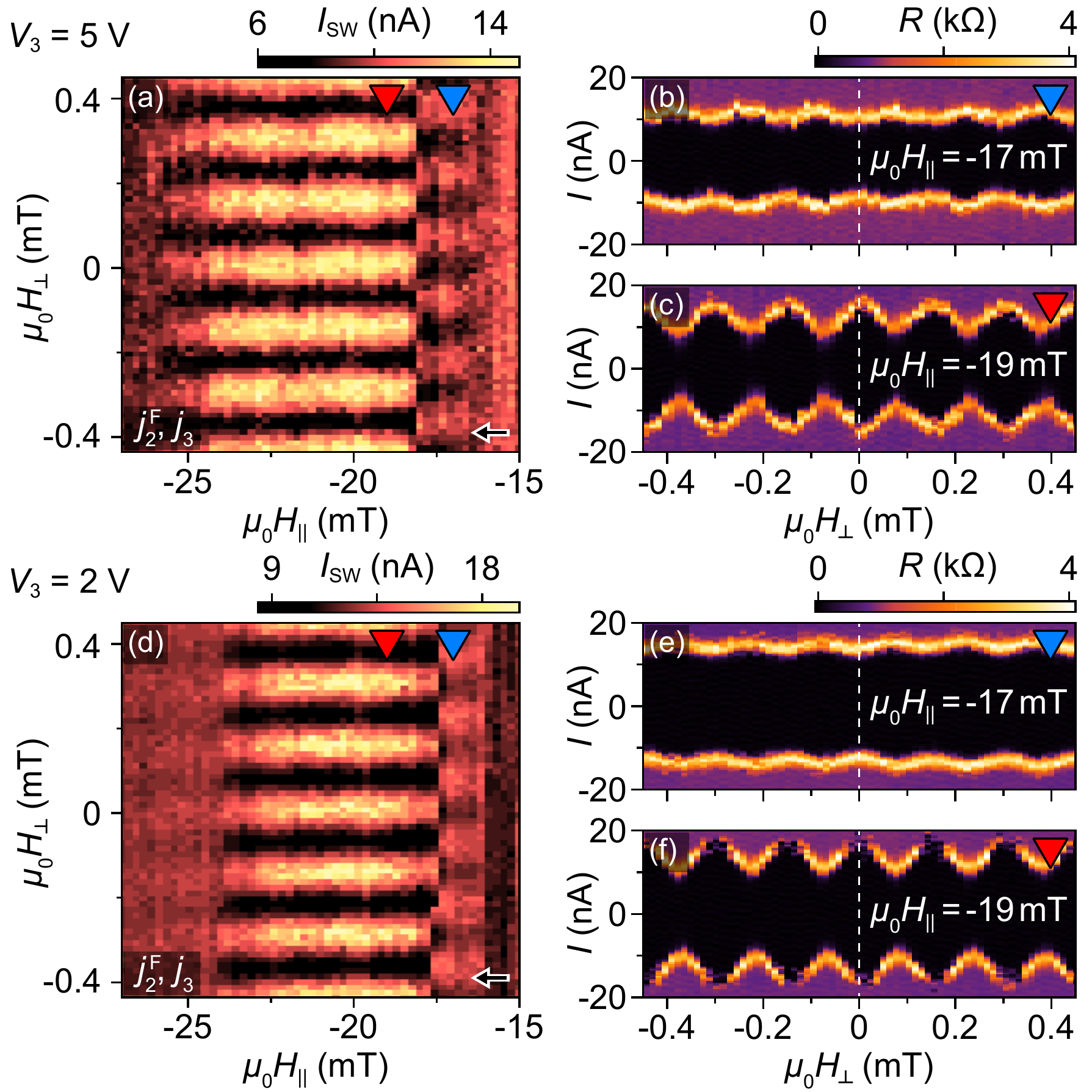}
\caption{\label{fig:S3}
Similar to the main-text Fig. 3 but measured at (a)-(c) $V_3 = 5$~V and (d)-(f) $V_3 = 2$~V, showing qualitatively identical $0\,--\,\pi$ transition features.
The variation in the transition-field values is due to the magnetic noise from the stochastic domain switching as discussed in the main text.
}
\end{minipage}\qquad
\begin{minipage}[t][-5ex][b]{.47\textwidth}
\includegraphics[width=\linewidth]{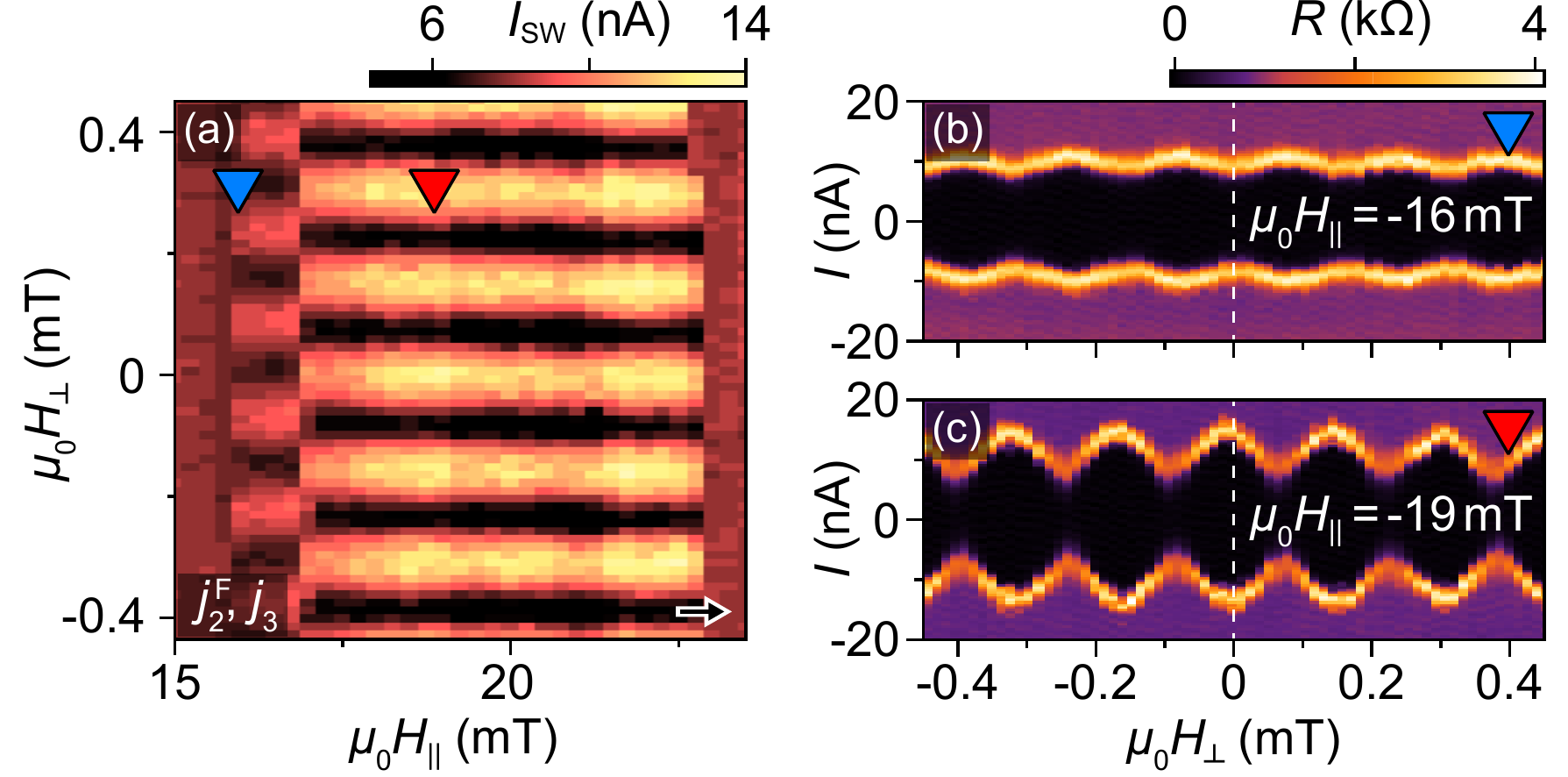}
\caption{\label{fig:S4}
Similar to the main-text Fig.~3 but measured at increasing $H_\parallel$ after polarizing the wire at $\mu_0 H_\parallel = -100$~mT, showing qualitatively identical $j_2^{\rm F}$\,-\,$j_3$ interferometer behavior.
}
\end{minipage}
\end{figure*}

\begin{figure*}
\centering
\begin{minipage}[t][15ex][b]{.47\textwidth}
\includegraphics[width=\linewidth]{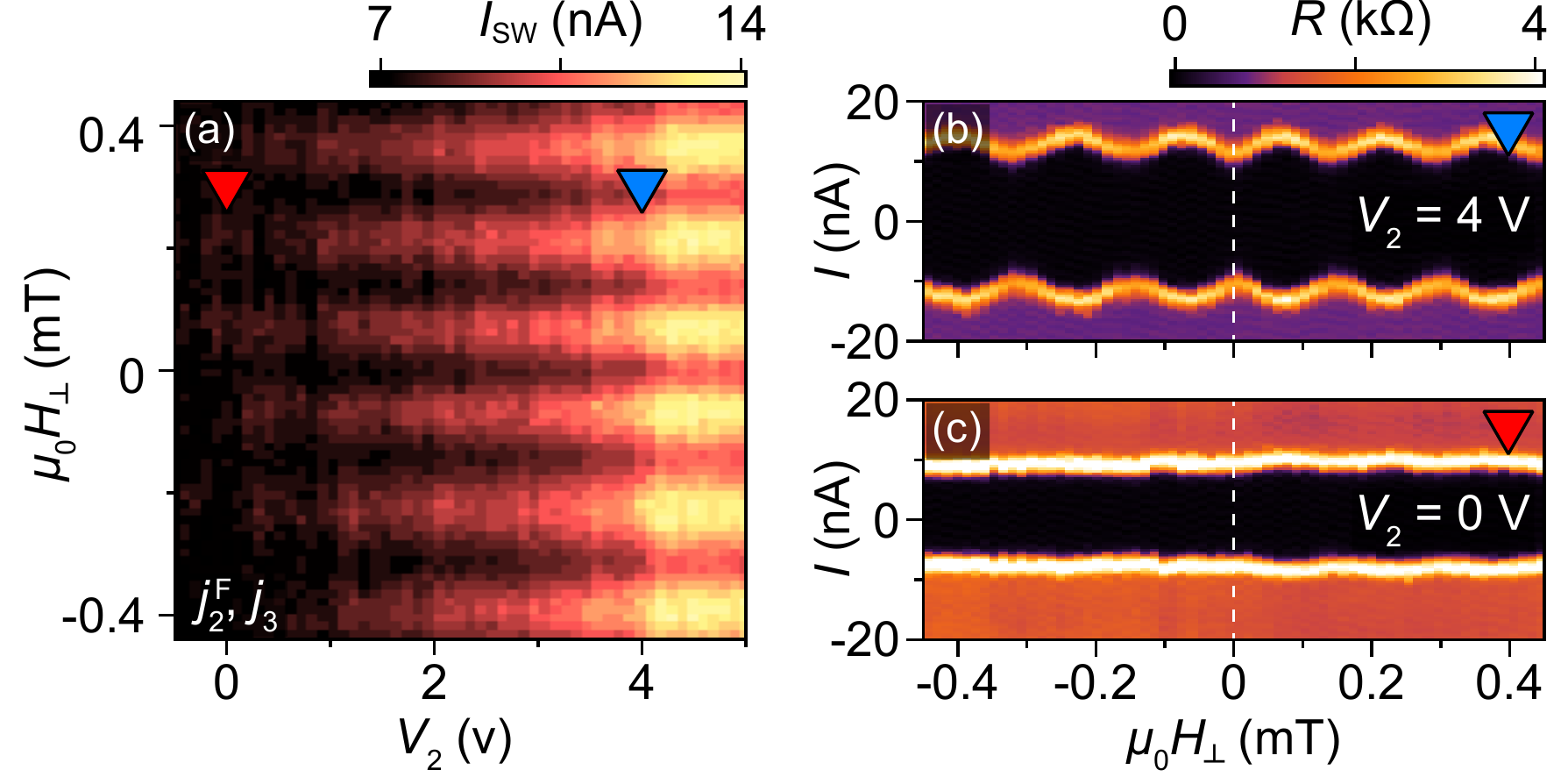}
\caption{\label{fig:S5}
(a) Switching current, $I_{\rm SW}$, of the $j_2^{\rm F}$\,-\,$j_3$ interferometer in the $\pi$ phase as a function of flux-threading magnetic field, $H_\perp$, and gate voltage, $V_2$.
The $\pi$ phase persists throughout the full range, until the supercurrent through $j_2^{\rm F}$ is fully suppressed.
The data were taken at $H_\parallel = 0$, after demagnetizing the wire at $H_\parallel^{\rm D}=-23$~mT (see main-text Fig.~5) and junction-gate voltages $V_1 = -6$~V, $V_3 = 0.5$~V, and $V_4 = 0$ (after the electrostatic discharge damage of $j_4$).
(b) and (c) Current-phase relations measured at $V_2 = 4$ and $0$~V, respectively, showing $\pi$-junction behavior.
}
\end{minipage}\qquad
\begin{minipage}[t]{.47\textwidth}
\includegraphics[width=\linewidth]{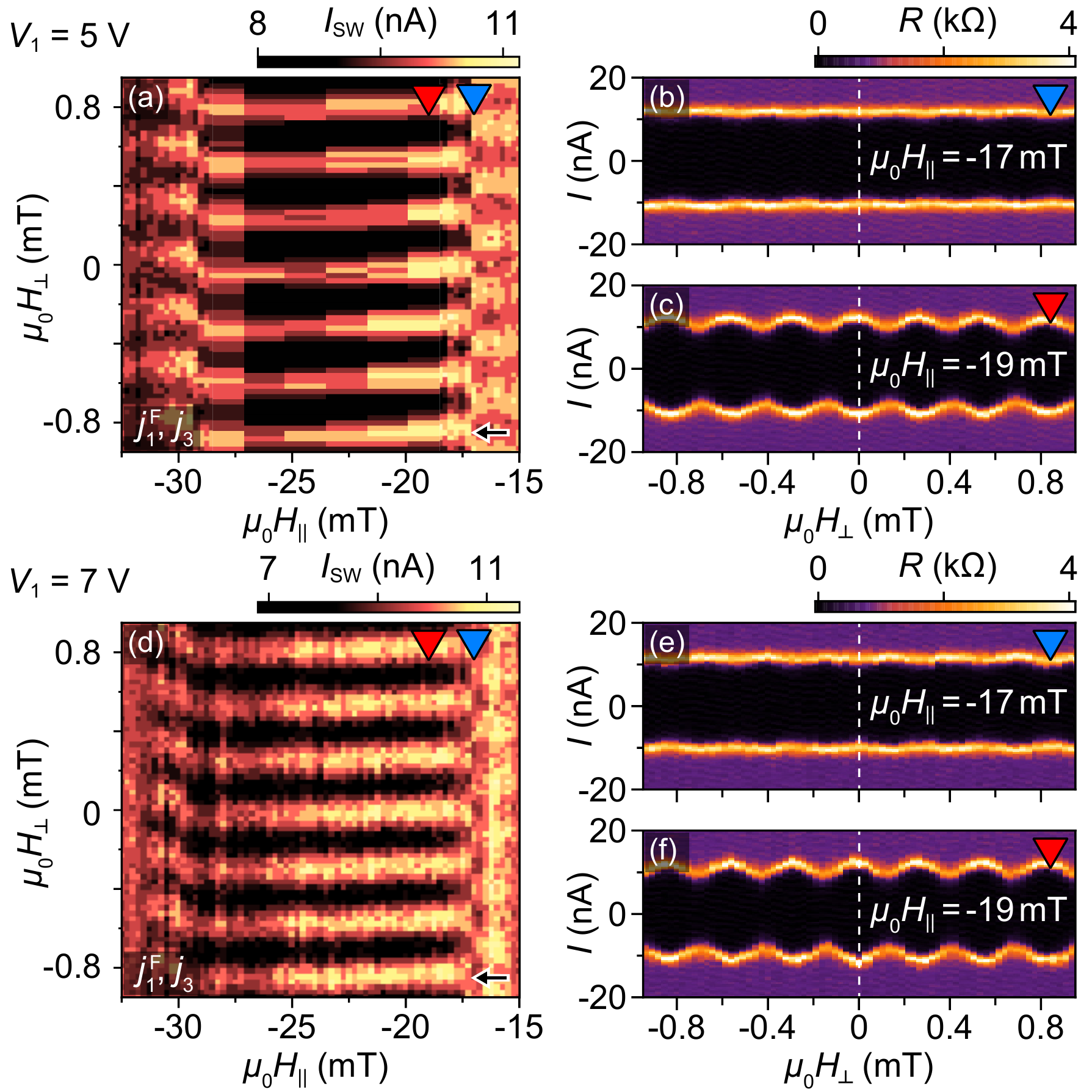}
\caption{\label{fig:S6}
Similar to the main-text Fig. 3 but measured for $j_1^{\rm F}$\,-\,$j_3$ interferometer at (a)-(c) $V_1 = 5$~V and (d)-(f) $V_3 = 7$~V, displaying the characteristic $\pi$\,--\,0 transition at the onset of the superconducting window, qualitatively identical to the $j_2^{\rm F}$\,-\,$j_3$ interferometer.
The CPR displays a small drift with $H_\parallel$, presumably due to a slight tilt of the chip, and a more abrupt shift of $\sim\pi$ at the end of the superconducting window.
The data were taken at junction-gate voltages $V_2 = -6$~V, $V_3 = 0.25$~V, and $V_4 = 0$ (after the electrostatic discharge damage of $j_4$).
}
\end{minipage}
\end{figure*}

\begin{figure*}
\centering
\begin{minipage}[t]{.47\textwidth}
\includegraphics[width=\linewidth]{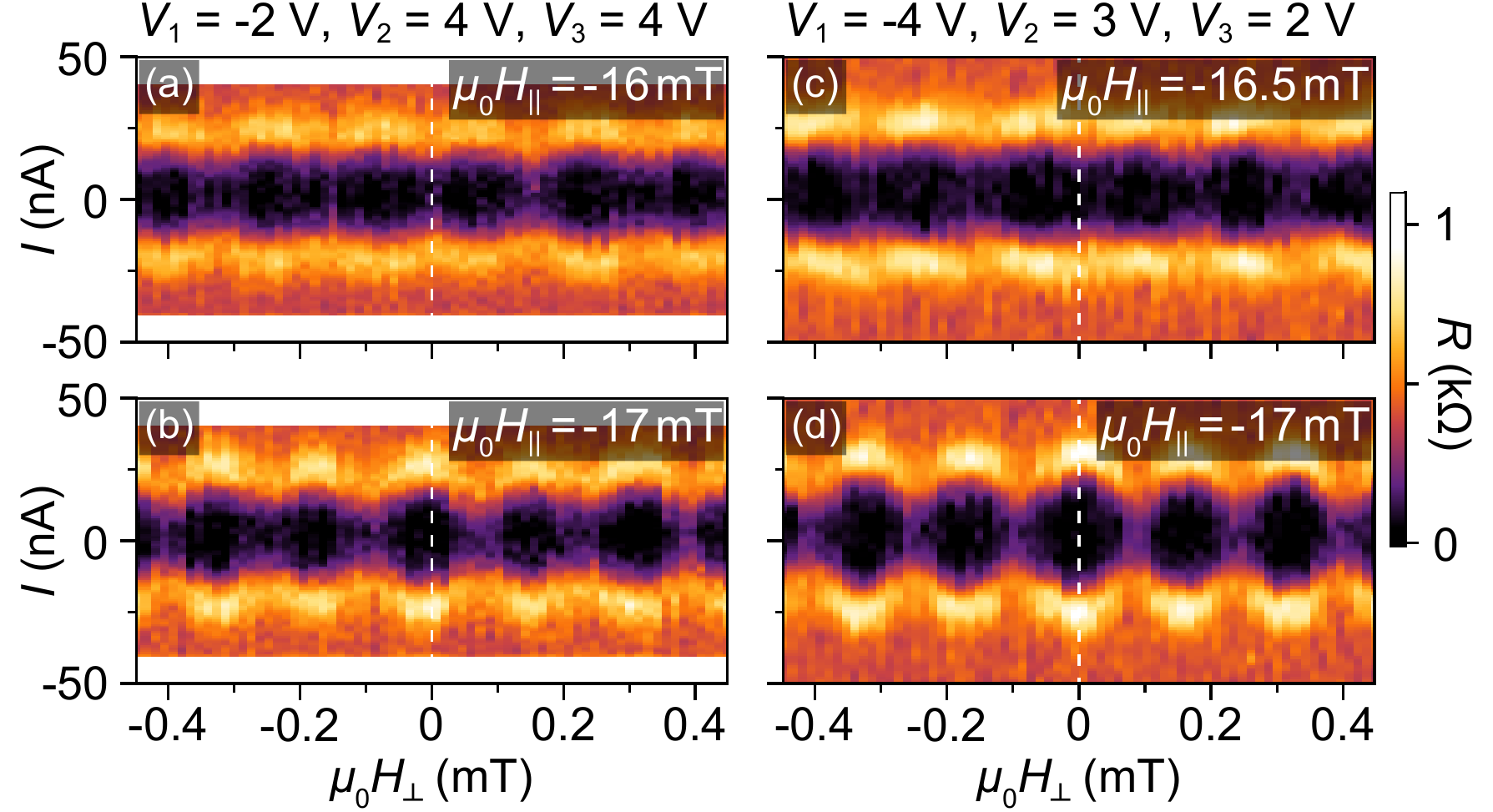}
\caption{\label{fig:S7}
Representative measurements of the second multi-interferometer device in $j_2^{\rm F}$\,-\,$j_3$.
Differential resistance, $R$, as a function of current bias, $I$, and flux-threading perpendicular magnetic field, $H_\perp$, measured at decreasing parallel external magnetic field, $H_\parallel$, after polarizing the wire at $\mu_0 H_\parallel = 100$~mT.
The data were taken at two different junction-gate voltage configurations: (a)-(b) $V_1 = -2$~V, $V_2 = 4$~V, $V_3 = 4$, and (c)-(d) $V_1 = -4$~V, $V_2 = 3$~V, $V_3 = 2$.
In both gate configurations, the relative phase of the switching current oscillations abruptly shift from $\pi$ to 0 around $\mu_0 H_\parallel = -17$~mT..
}
\end{minipage}\qquad
\begin{minipage}[t]{.47\textwidth}
\includegraphics[width=\linewidth]{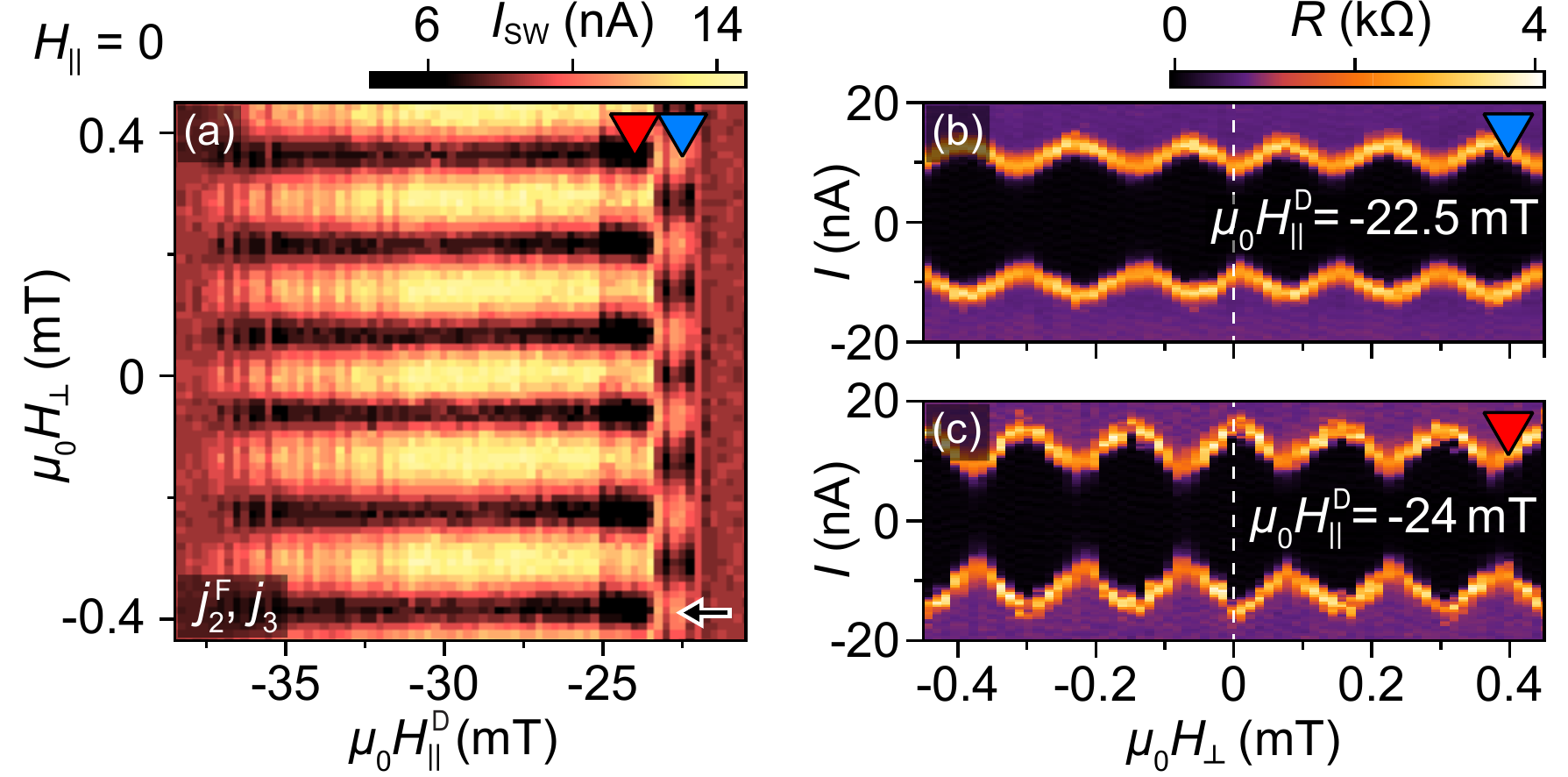}
\caption{\label{fig:S8}
Similar to the main-text Fig. 3 but measured at $H_\parallel = 0$, after demagnetizing the wire at variable $H_\parallel^{\rm D}$ (see main-text Fig.~5).
The relative phase of the switching current oscillations abruptly shift from $\pi$ to 0 around $\mu_0 H_\parallel^{\rm D} = -23.5$~mT.
}
\end{minipage}
\end{figure*}

\begin{figure*}
\centering
\begin{minipage}[t]{.47\textwidth}
\includegraphics[width=\linewidth]{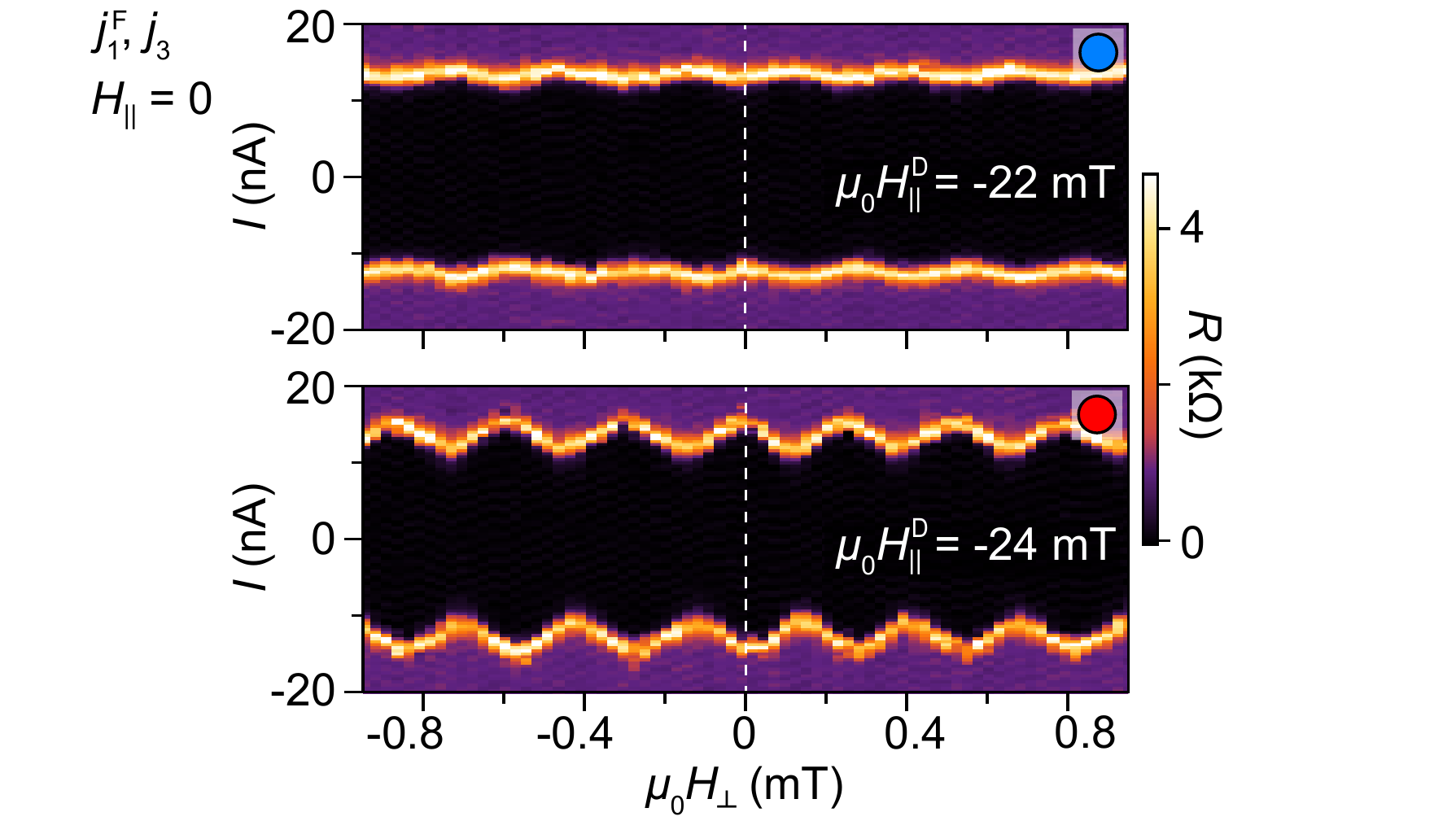}
\caption{\label{fig:S9}
Similar to the main-text Fig.~5 but measured for $j_1^{\rm F}$\,-\,$j_3$ interferometer.
(a) Differential resistance, $R$, as a function of current bias, $I$, and flux-threading magnetic field, $H_\perp$, measure at zero parallel external field ($H_\parallel = 0$) showing a $\pi$-shifted current-phase relation.
The data were taken after polarizing and demagnetizing the wire at $\mu_0 H_\parallel^{\rm D}=-22$~mT.
(b) Similar to (a) but taken after demagnetizing the wire at $\mu_0 H_\parallel^{\rm D}=-24$~mT, showing $0$-junction behavior.
}
\end{minipage}\qquad
\begin{minipage}[t]{.47\textwidth}
\includegraphics[width=\linewidth]{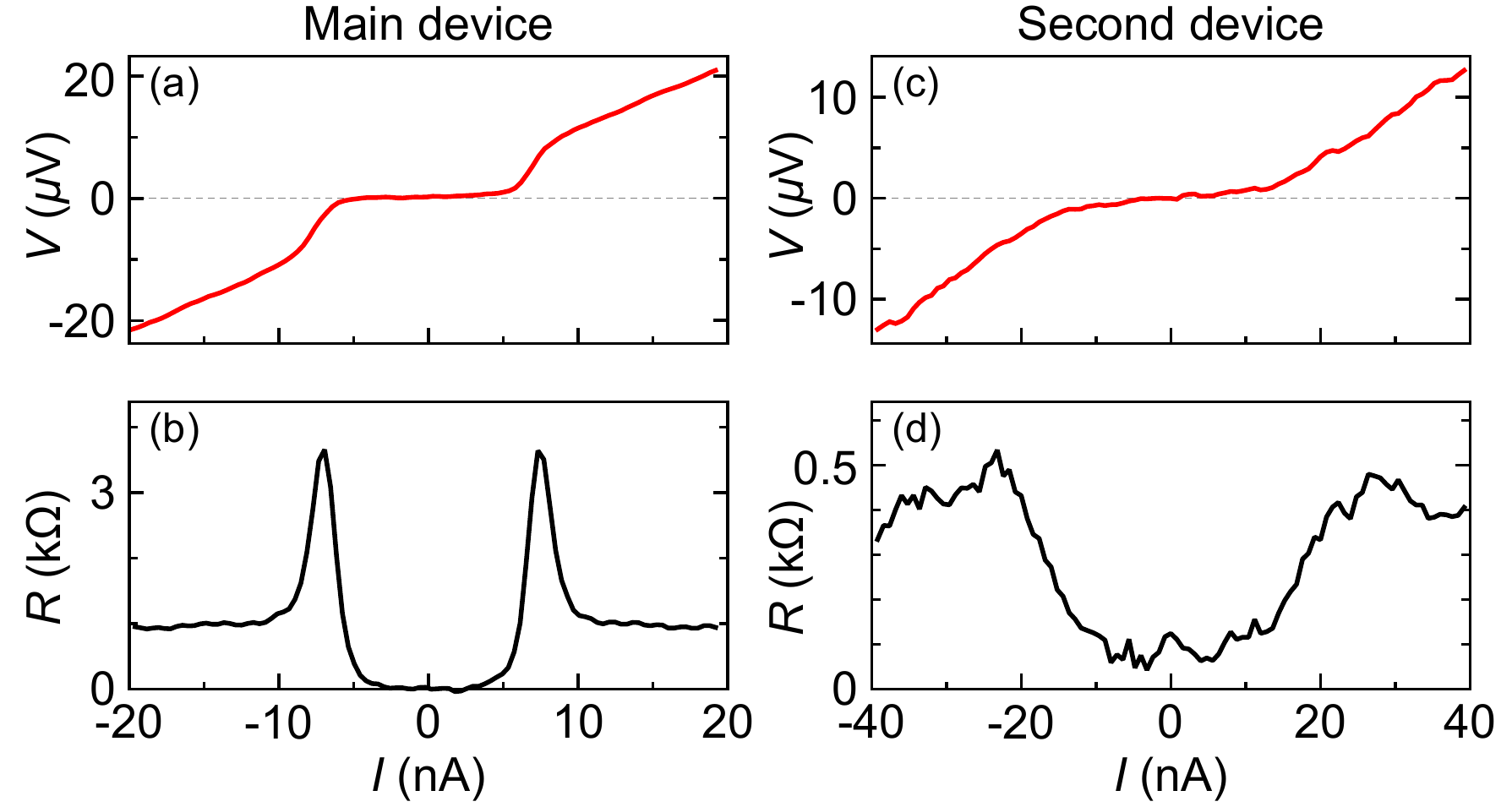}
\caption{\label{fig:S10}
Four-terminal dc voltage, $V$, and differential resistance, $R$, as a function of current bias, $I$, measured for (a)-(b) the main and (c)-(d) the second mulit-interferometer devices after zero-field cooling, at zero junction-gate voltages, before applying external magnetic field.
}
\end{minipage}
\end{figure*}

\begin{figure*}[t!]
\includegraphics[width=0.5\linewidth]{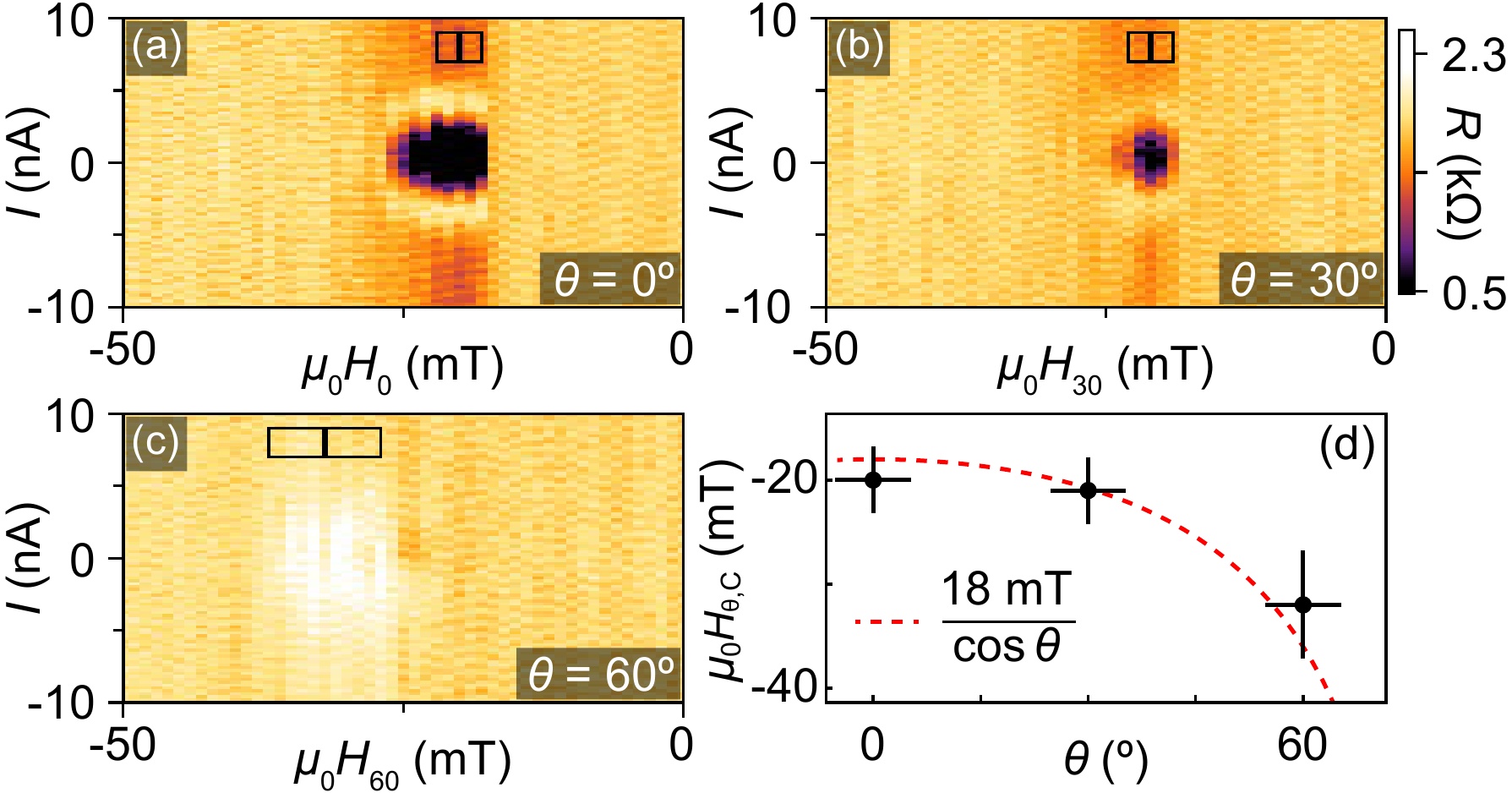}
\caption{\label{fig:S11}
(a)--(c) Differential resistance, $R$, as a function of current bias, $I$, and in-plane external magnetic field, $H_\theta$, measured for the $j_2^{\rm F}$ junction of the main multi-interferometer sweeping $H_\theta$ from positive negative values at $\theta$ = (a) 0, (b) 30, and (c) 60 degree.
(d) The measured coercive fields, $H_{\theta, \rm C}$, as indicated for each sweep in (a)--(c), decreases as 1/cos$(\theta)$. The error bars illustrate the uncertainty coming from the actual wire angle with respect to the vector-magnet axes and the superconducting feature width.
}
\end{figure*}

\begin{figure*}[t!]
\includegraphics[width=\linewidth]{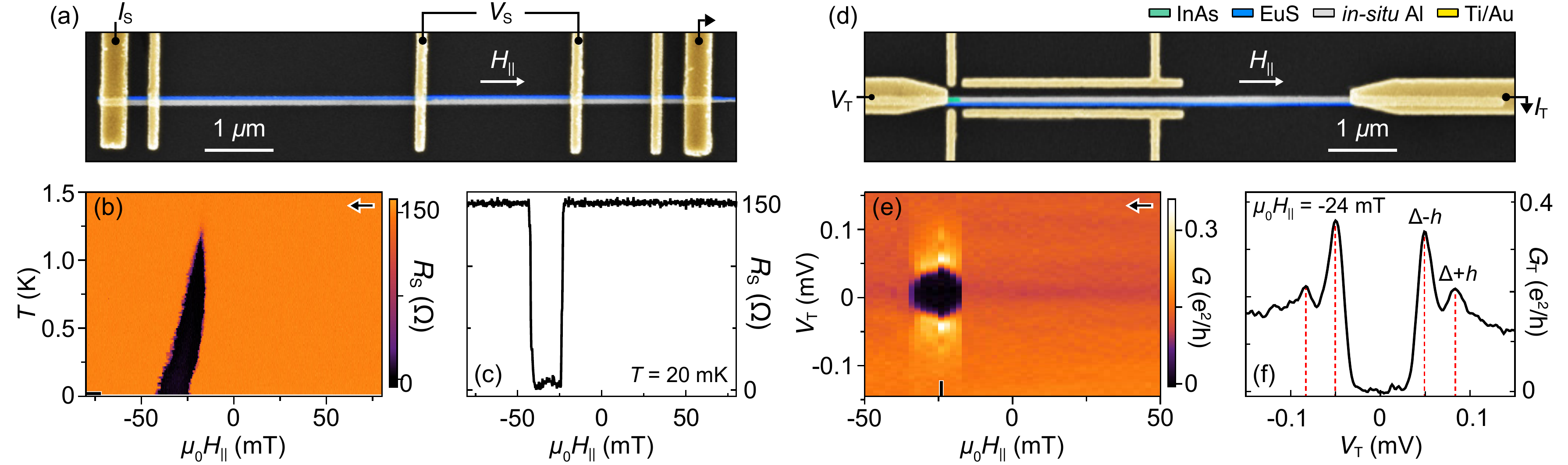}
\caption{\label{fig:S12}
(a) Color-enhanced micrograph of the additional shell-characterization device.
(b) Four-probe differential resistance of Al/EuS shell, $R_{\rm S} = {\rm d}V_{\rm S}/{\rm d}I_{\rm S}$, as a function of temperature, $T$, and parallel external magnetic field $\mu_0H_\parallel$, shows a skewed superconducting phase diagram with a critical critical temperature $T_{\rm C0}=(1.2\pm0.1)$~K around the coercive field.
(c) Line-cut taken from (b) at $T=20$~mK showing a superconducting window away from $H_\parallel = 0$ and the normal-state resistance $R_{\rm N} = 150~\Omega$.
(d) Color-enhanced micrograph of the additional tunneling-spectroscopy device.
(e) Differential tunneling conductance, $G_{\rm T} = {\rm d}I_{\rm T}/{\rm d}V_{\rm T}$, as a function of source-drain voltage bias, $V_{\rm T}$, and parallel external magnetic field $\mu_0H_\parallel$, shows a superconducting window away from $H_\parallel = 0$.
(f) Line-cut from (e) at $\mu_0H_\parallel=-24$~mT showing a superconducting gap with spin-split coherence peaks.}
\end{figure*}

\begin{figure*}
\centering
\begin{minipage}{.47\textwidth}
\includegraphics[width=\linewidth]{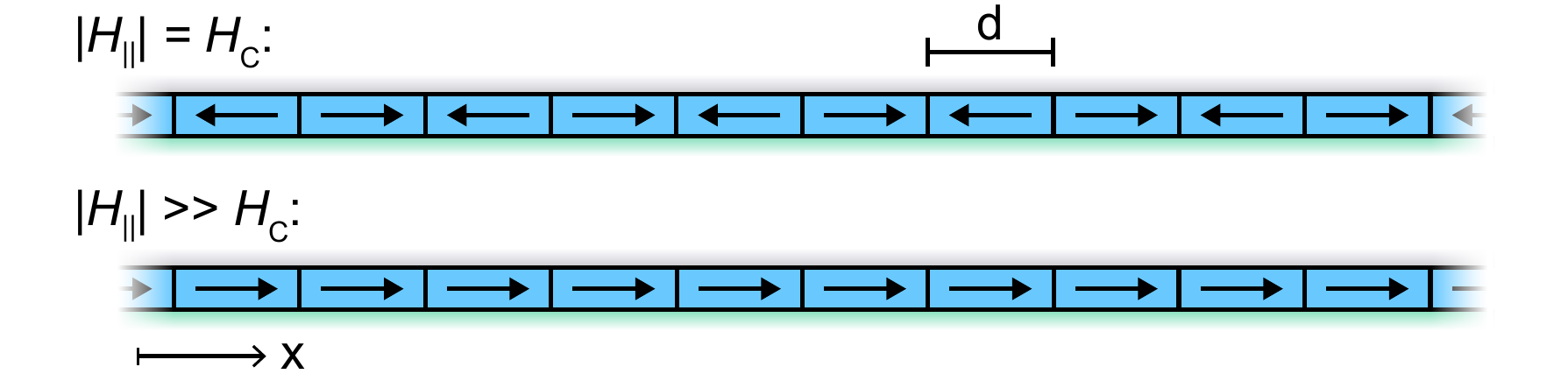}
\caption{\label{theory_S1} 
Sketch of the domain polarization in the magnetic shell at the coercive field, $\vert H_\parallel\vert = H_{\rm C}$, giving a randomized magnetization configuration and much larger fields, $\vert H_\parallel\vert = H_{\rm C}$, giving a fully polarized configuration.}
\end{minipage}\qquad
\begin{minipage}[t]{.47\textwidth}\includegraphics[width=\linewidth]{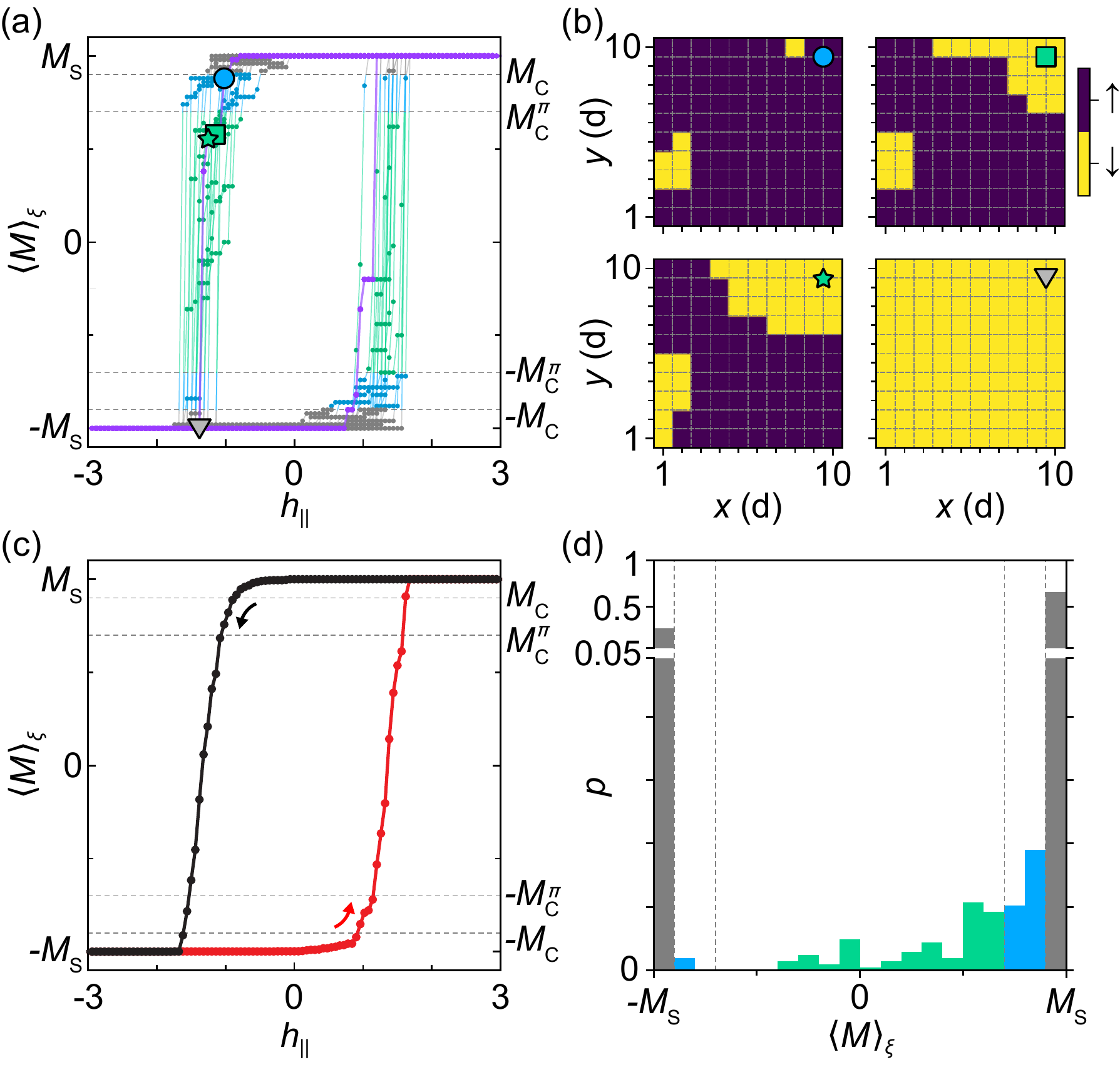}
\caption{\label{theory_S2} 
(a) Average magnetization, $\left< M \right>_\xi$, as a function of external magnetic field, $h_\parallel$, calculated using kinetic random-field Ising model [see Eq.~\eqref{eq:krfim_hamiltonian}] for 20 disorder realizations of a grid of $10 \times 10$ magnetic grains.
$M_{\rm S}$ is magnetization saturation value.
$M_{\rm C}$ and $M_{\rm C}^\pi$ are arbitrarily-chosen critical-magnetization values.
(b) Four magnetic-domain configurations of the grid taken from a selected down-sweep realization highlighted in (a) at equidistantly spaced $h_\parallel$ values.
(c) Disorder-averaged $\left< M \right>_\xi$ as a function of $h_\parallel$, deduced from the 20 realizations shown in (a).
(d) Histogram of the 20 down-sweep traces in (a) representing the probability, $p$, for the grid to have a particular $\left< M \right>_\xi$~value.}
\end{minipage}
\end{figure*}

\begin{figure*}
\centering
\begin{minipage}[t]{.47\textwidth}
\includegraphics[width=\linewidth]{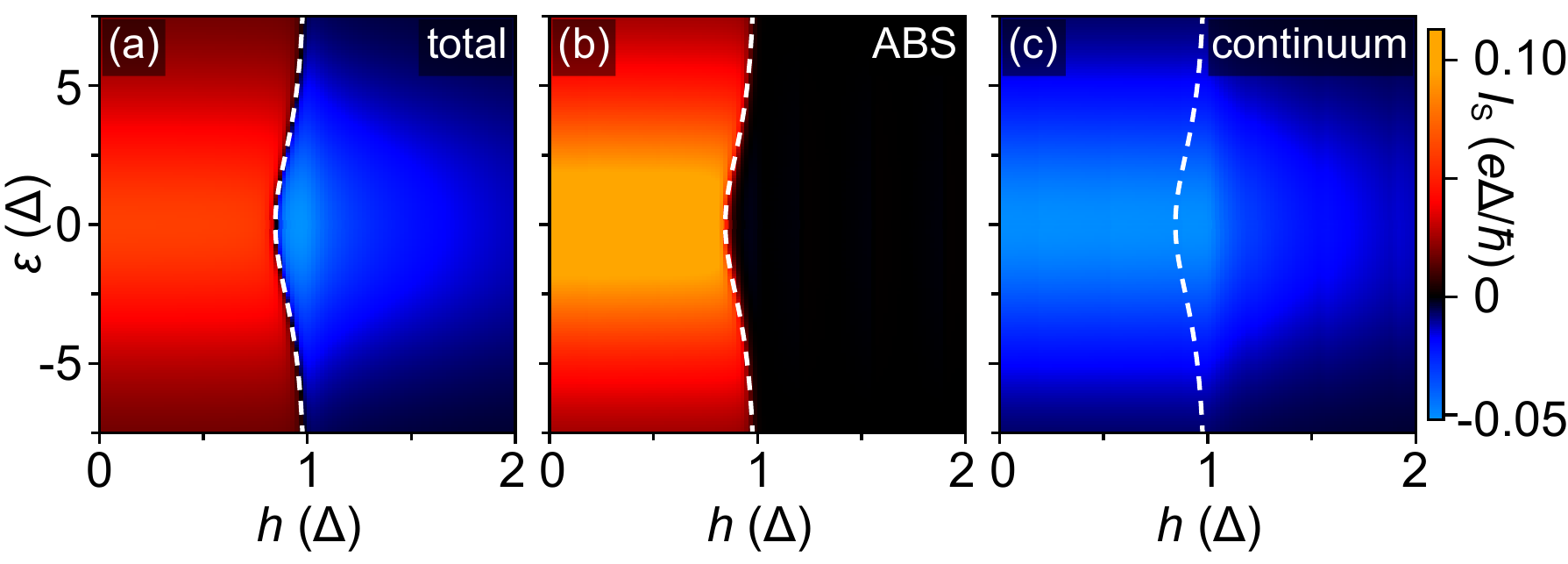}
\caption{\label{theory_S3}
(a) Calculated supercurrent, $I_{\rm S}$, through the hybrid junction in the single-barrier regime ($\Gamma_{\rm L} = 20 \Gamma_{\rm R} = 4\Delta$, same as in the main-text Fig.~4), as a function of the normal level energy, $\varepsilon$, and the homogeneous spin-splitting field, $h$, at the superconducting phase difference $\varphi=\pi/2$.
Panels (b) and (c) show $I_{\rm S}$ contributions from the Andreev bound states (ABSs) and the continuum of states.
White dashed lines correspond to Eq.~\eqref{Eq:solDet}.}
\end{minipage}\qquad
\begin{minipage}[t]{.47\textwidth}
\includegraphics[width=\linewidth]{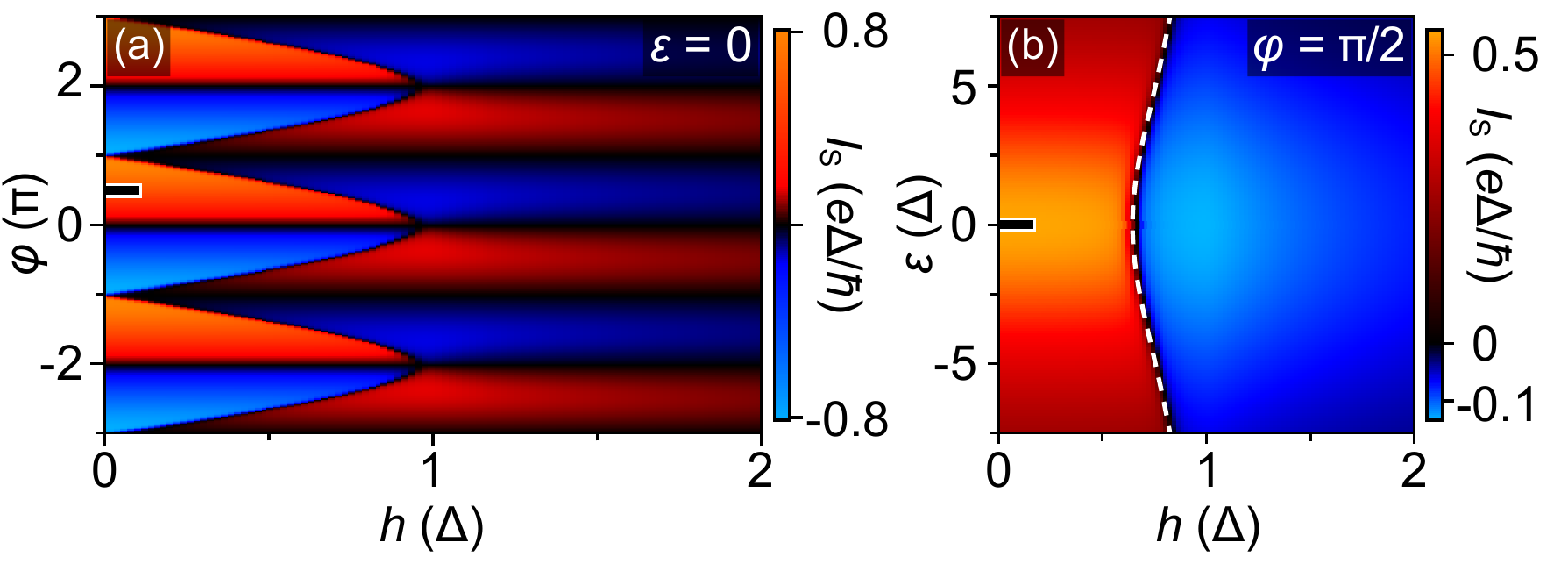}
\caption{\label{theory_S4}
(a) Calculated supercurrent, $I_{\rm S}$, through the hybrid junction in the open regime ($\Gamma_L=\Gamma_R=4\Delta$), as a function of the superconducting phase difference, $\varphi$, and the homogeneous spin-splitting field, $h$, at normal level energy $\varepsilon=0$.
(b) $I_{\rm S}$ as a function of $\varepsilon$ and $h$ for $\varphi=\pi/2$.
White dashed lines correspond to Eq.~\eqref{Eq:solDet}.}
\end{minipage}
\end{figure*}

\begin{figure*}
\centering
\begin{minipage}[t]{.47\textwidth}
\includegraphics[width=\linewidth]{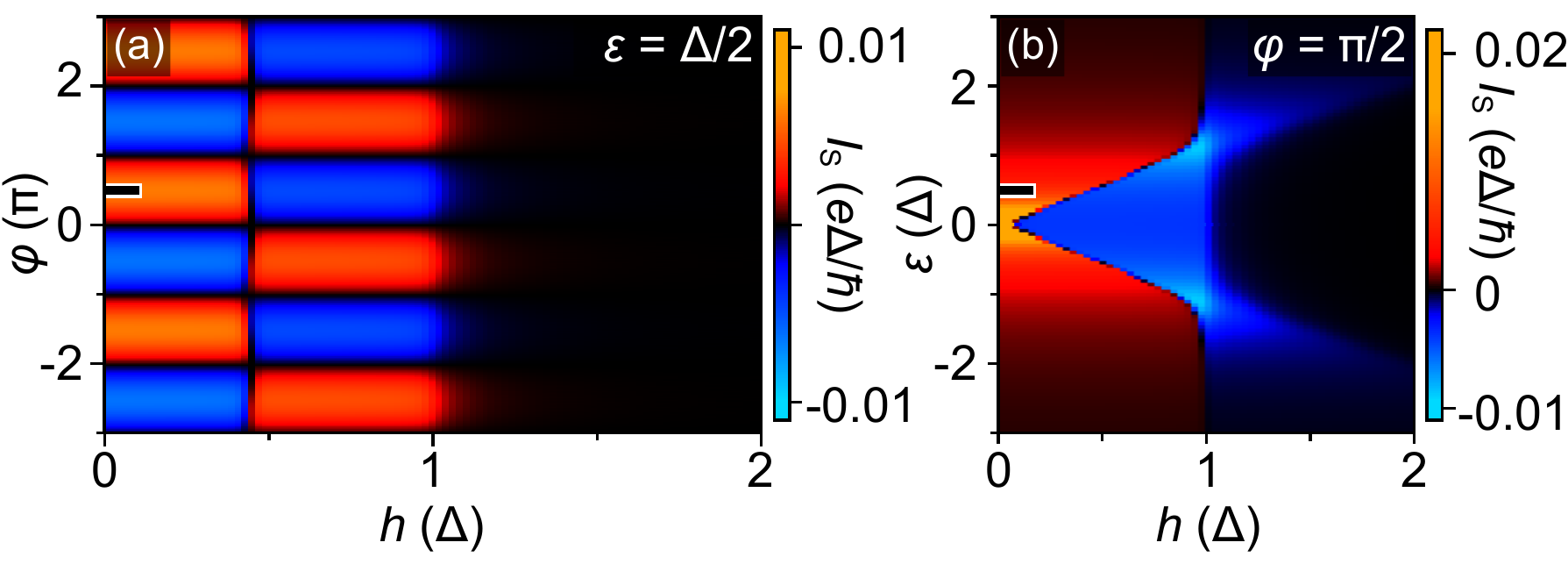}
\caption{\label{theory_S5}
(a) Calculated supercurrent, $I_{\rm S}$, through the hybrid junction in the quantum-dot regime ($\Gamma_L=\Gamma_R=\Delta/20$), as a function of the superconducting phase difference, $\varphi$, and the homogeneous spin-splitting field, $h$, at normal level energy $\varepsilon=\Delta/2$.
(b) $I_{\rm S}$ as a function of $\varepsilon$ and $h$ for $\varphi=\pi/2$.
White dashed lines correspond to Eq.~\eqref{Eq:solDet}.}
\end{minipage}\qquad
\begin{minipage}[t]{.47\textwidth}
\includegraphics[width=\linewidth]{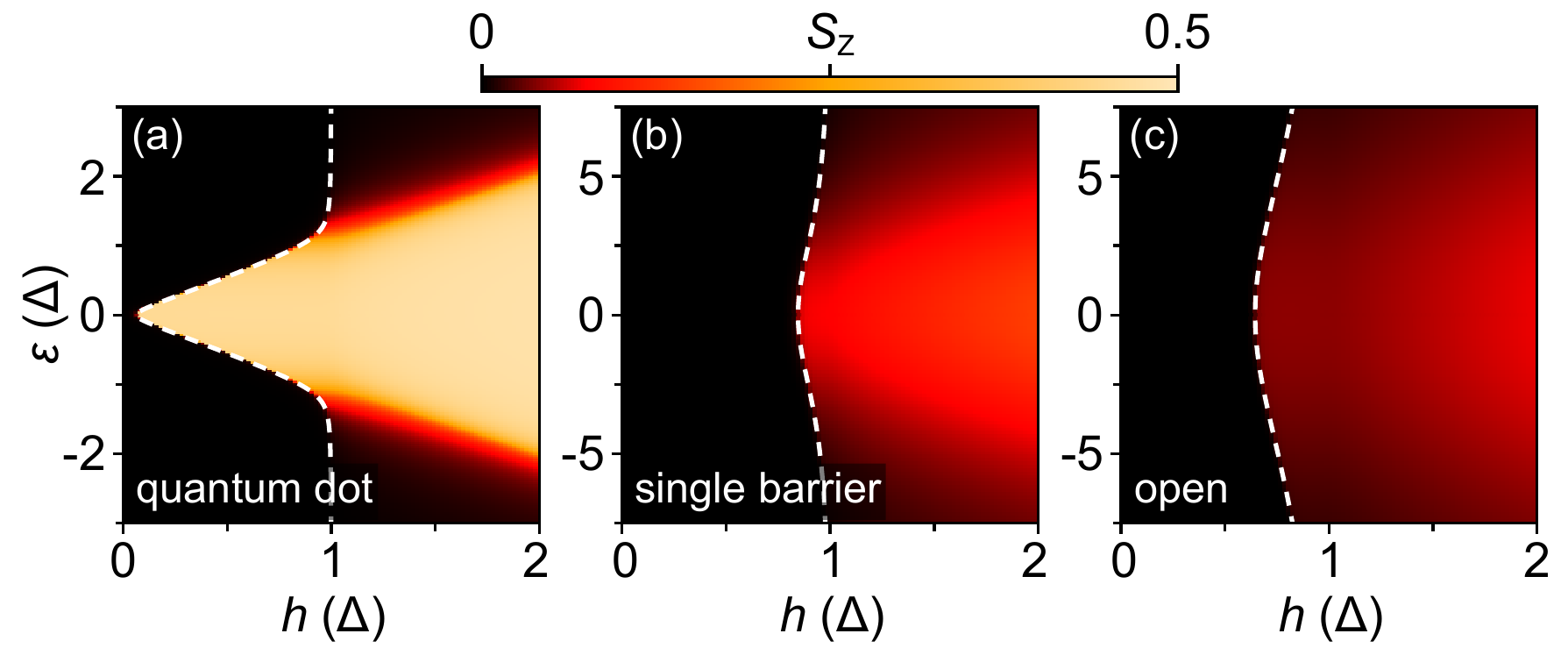}
\caption{\label{theory_S6}
Calculated spin polarization of the hybrid junction, $S_{\rm Z}$, as a function of the normal level energy, $\varepsilon$, and the homogeneous spin-splitting field, $h$, in the quantum-dot (a), single-barrier (b), and open (c) regimes.
Parameters are the same as in Figs.~\ref{theory_S3}--\ref{theory_S5}.
White dashed lines correspond to Eq.~\eqref{Eq:solDet}.}
\end{minipage}
\end{figure*}

\begin{figure*}
\centering
\begin{minipage}{.47\textwidth}
\includegraphics[width=\linewidth]{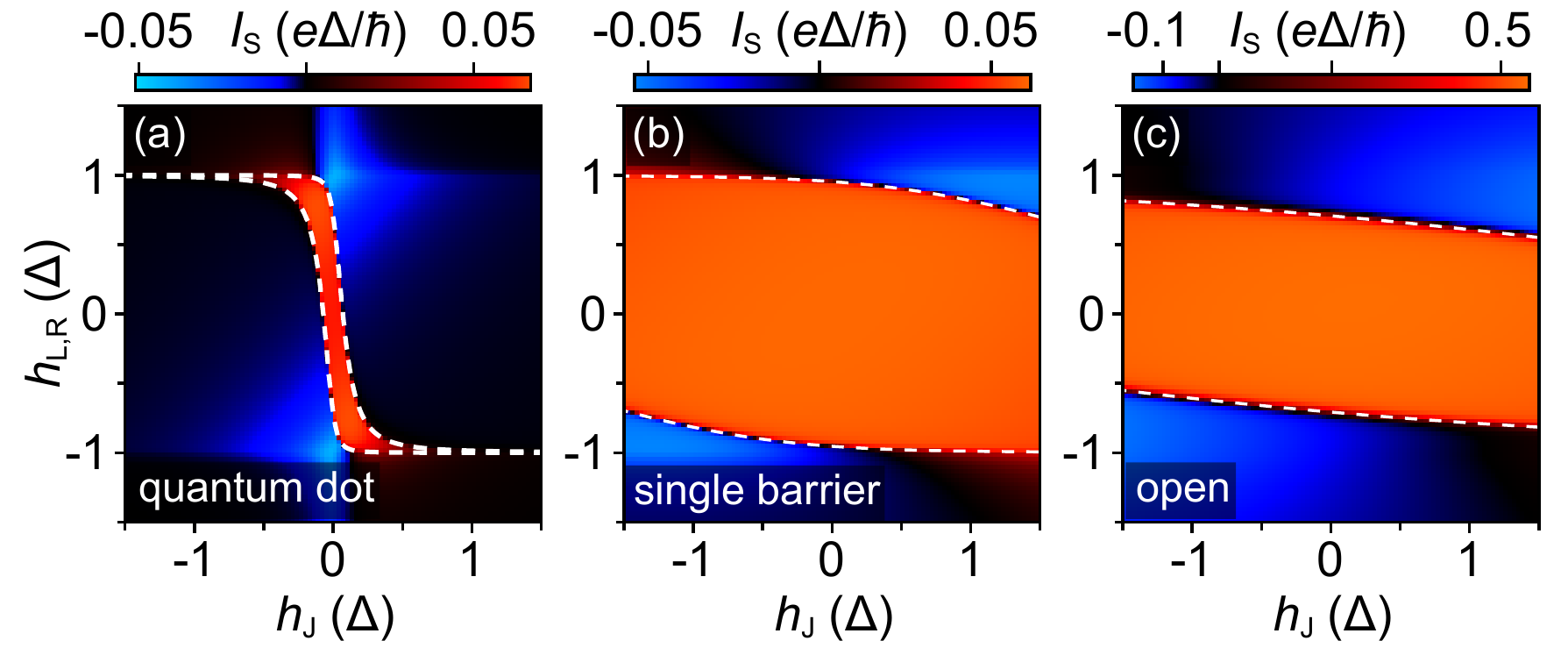}
\caption{Calculated supercurrent, $I_{\rm S}$, as a function of the spin-splitting field in the junction, $h_{\rm J}$, and the leads, $h_{\rm L,R}$, in the quantum-dot (a), single-barrier (b), and open (c) regimes at $\varphi=\pi/2$ and $\varepsilon=0$.
Parameters are the same as in Figs.~\ref{theory_S3}--\ref{theory_S6}.
White dashed lines correspond to Eq.~\eqref{Eq:solDet}.
\label{theory_S7}}
\end{minipage}\qquad
\begin{minipage}[t]{.47\textwidth}
\includegraphics[width=\linewidth]{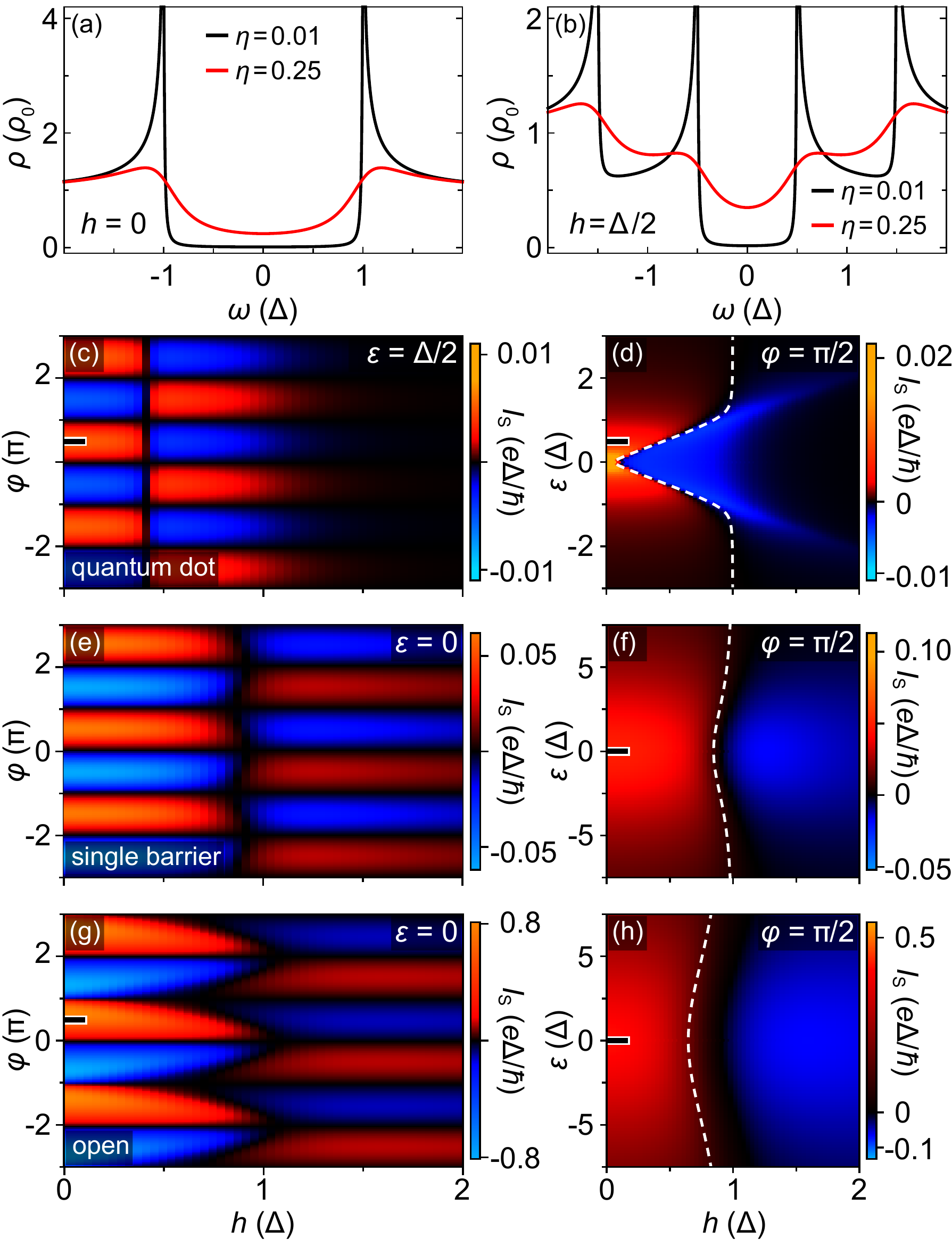}
\caption{
(a) and (b) Calculated superconducting density of states, $\rho$, in units of the normal state conductance, $\rho_0$, as a function of the electron energy $\omega$,  for two different Dynes parameter values ($\eta$ = 0.01 and 0.25) at the homogeneous spin-splitting field $h=0$ (a) and $h=\Delta/2$ (b).
The superconducting gap is softened significantly for $\eta = 0.25$ compared to $\eta = 0.01$.
(c) Calculated supercurrent, $I_{\rm S}$, through the hybrid junction in the quantum-dot regime for $\eta = 0.25$, as a function of the superconducting phase difference, $\varphi$, and $h$.
(d) $I_{\rm S}$ as a function of normal level energy, $\varepsilon$, and $h$.
(e) and (f) Similar to (c) and (d) but in the single-barrier regime.
(g) and (h) Similar to (c) and (d) but in the open regime.
Other parameters are the same as in Figs.~\ref{theory_S3}--\ref{theory_S6}.
White dashed lines correspond to Eq.~\eqref{Eq:solDet}.
\label{theory_S8}}
\end{minipage}
\end{figure*}

\bibliography{bibfile}